\documentclass[twocolumn]{aastex631}
\usepackage{xcolor}

\begin{document}

\newcommand{\kms}{km~s$^{-1}$}	\newcommand{\cms}{cm~s$^{-2}$}
\newcommand{\msun}{M$_{\odot}$} \newcommand{\rsun}{$R_{\odot}$} 
\newcommand{\teff}{$T_{\rm eff}$}
\newcommand{\mas}{mas~yr$^{-1}$}
\newcommand{\logg} {\log \textsl{\textrm{g}}}
\newcommand{\simon}[1]{\textcolor{teal}{#1}}

\title{White Dwarf Merger Remnants: The DAQ Subclass}

\author[0000-0001-6098-2235]{Mukremin Kilic} 
\affiliation{Homer L. Dodge Department of Physics and Astronomy, University of Oklahoma, 440 W. Brooks St., Norman, OK, 73019 USA}

\author[0000-0003-2368-345X]{Pierre Bergeron} 
\affiliation{D\'epartement de Physique, Universit\'e de Montr\'eal, C.P. 6128, Succ. Centre-Ville, Montr\'eal, QC H3C 3J7, Canada}

\author[0000-0002-9632-1436]{Simon Blouin}
\affiliation{Department of Physics and Astronomy, University of Victoria, Victoria BC V8W 2Y2, Canada}

\author[0009-0009-9105-7865]{Gracyn Jewett}  
\affiliation{Homer L. Dodge Department of Physics and Astronomy, University of Oklahoma, 440 W. Brooks St., Norman, OK, 73019 USA}

\author[0000-0002-4462-2341]{Warren R.\ Brown}
 \affiliation{Center for Astrophysics, Harvard \& Smithsonian, 60 Garden Street, Cambridge, MA 02138 USA}

\author[0000-0001-7143-0890]{Adam Moss}  
\affiliation{Homer L. Dodge Department of Physics and Astronomy, University of Oklahoma, 440 W. Brooks St., Norman, OK, 73019 USA}


\begin{abstract}

Four years after the discovery of a unique DAQ white dwarf with a hydrogen-dominated and carbon-rich atmosphere, we report the discovery of four new DAQ white dwarfs, including two that were not recognized properly in the literature. We find all five DAQs in a relatively narrow mass and temperature range of $M=1.14-1.19$ \msun\ and $T_{\rm eff}=13,000-17,000$ K. In addition, at least two show photometric variations due to rapid rotation with $\approx10$ min periods. All five are also kinematically old, but appear photometrically young with estimated cooling ages of about 1 Gyr based on standard cooling tracks, and their masses are roughly twice the mass of the most common white dwarfs in the solar neighborhood. These characteristics are smoking gun signatures of white dwarf merger remnants. Comparing the DAQ sample with warm DQ white dwarfs, we demonstrate that there is a range of hydrogen abundances among the warm DQ population, and the distinction between DAQ and warm DQ white dwarfs is superficial. We discuss the potential evolutionary channels for the emergence of the DAQ subclass, and suggest that DAQ white dwarfs are trapped on the crystallization sequence, and may remain there for a significant fraction of the Hubble time. 

\end{abstract}


\section{INTRODUCTION}

Short period double white dwarfs lose angular momentum through gravitational wave radiation and merge to create a variety of
interesting phenomena, including Type Ia supernovae \citep{webbink84,iben84}, hot subdwarfs \citep{heber09}, R Coronae Borealis
stars \citep{clayton07}, ultramassive white dwarfs, or collapse into neutron stars \citep{nomoto85}. Binary population synthesis
models demonstrate that the majority of double white dwarf mergers have a combined mass below the Chandrasekhar limit
\citep{toonen12}. For reference, the field white dwarf mass distribution shows a dominant peak at 0.6 \msun\
\citep{kilic20,mccleery20}. Hence, mergers of the most common white dwarfs should form ultramassive white dwarfs with
$M\approx1.2$ \msun\ \citep[e.g.,][]{kawka23}.

For the non-explosive CO + CO white dwarf mergers, the merger remnants experience a luminous giant phase for about ten thousand
years, after which they evolve into single massive white dwarfs with typical rotation periods of 10-20 min \citep[][see also \citealt{yoon07,loren09,shen12}]{schwab21}. Such rotation rates are
significantly shorter than the day-long rotation periods seen in pulsating white dwarfs \citep{kawaler15,hermes17b}, but are now routinely
observed in ultramassive and/or magnetic white dwarfs \citep{pshirkov20,kawka20,caiazzo21,kilic21b,kilic23a,williams22,moss23}.
  
Hot DQ white dwarfs\footnote{DQ white dwarfs are a spectral class of white dwarfs showing atomic or molecular carbon features \citep{sion83}.} with $T_{\rm eff} \approx$ 18,000-24,000 K \citep{dufour08} stand out among the population of rapidly rotating isolated white dwarfs. In addition to fast rotation, hot DQ white dwarfs have high masses $M\geq0.8$ \msun, high incidence of magnetism, and unusual kinematics for their age. Even though the emergence of cooler DQ white dwarfs below $T_{\rm eff} = 10,000$~K is well explained by the dredge-up of carbon in helium dominated atmospheres \citep{pelletier86,bedard22a}, the same dredge-up model fails to explain the chemical composition of warmer DQs \citep{coutu19,koester19}. Hence, hot DQs must follow a different evolutionary path. All evidence points to a merger origin \citep{dunlap15,coutu19,kawka23}.

\citet{hollands20} reported the discovery of a new spectral class, a DAQ white dwarf with a hydrogen-dominated and carbon-rich atmosphere (WD J055134.612+413531.09, hereafter J0551+4135).
Its unusual composition, large mass (1.14 \msun), and fast kinematics
strongly favor a white dwarf merger origin.  Large masses, carbon enriched atmospheres, and fast kinematics are also common characteristics of hot DQ white dwarfs, but it is not clear what causes the unusual atmospheric composition in J0551+4135 or the emergence of the DAQ subclass.

Here we report the discovery of four new DAQ white dwarfs, including three within 100 pc of the Sun. We describe our discovery observations of two new DAQ white dwarfs in Section 2, and present a detailed model atmosphere analysis in Section 3. We present the identification of two additional DAQ white dwarfs that were overlooked in the literature in Section 4, and discuss the distinction between DAQ and DQA white dwarfs in Section 5. We present the physical parameters of the DAQ sample and discuss possible evolutionary channels to explain their unusual compositions in Section 6. We conclude in Section 7.

\section{The Discovery of two New DAQ White Dwarfs}

As part of our efforts to constrain the physical properties of the massive white dwarfs in the solar neighborhood (Jewett et al., in prep.),
we obtained follow-up spectroscopy of $M\geq0.9$ \msun\ white dwarf candidates in the Montreal White Dwarf Database
100 pc sample \citep{dufour17}. To take advantage of the Pan-STARRS photometry in our atmospheric model fits, we further limited
our follow-up survey to the Pan-STARRS footprint. We also restricted our sample to objects with $T_{\rm eff}\geq11,000$ K so that
helium lines could be detected and used to constrain the atmospheric parameters. Our initial spectroscopic observations
at the Apache Point Observatory 3.5m telescope revealed two of these candidates, J0831$-$2231 and J2340$-$1819 (GD 1222),
as potential DAQs. Table \ref{tab1} presents the photometry and astrometry for these targets.

\begin{deluxetable*}{ccccc}
\tablecolumns{5} \tablewidth{0pt}
\tablecaption{New DAQ white dwarfs. \label{tab1}}
\tablehead{\colhead{Parameter} & \colhead{J0205+2057} & \colhead{J0831$-$2231} & \colhead{J0958+5853} & \colhead{J2340$-$1819}}
\startdata
Name & G35-26 & WDJ083135.57-223133.63 & SDSS J095837.00+585303.0 & GD 1222 \\
RA &  02:05:49.45 & 08:31:35.42 & 09:58:36.93 & 23:40:43.98 \\ 
DEC & +20:57:03.96 & $-$22:31:30.13 & +58:53:03.00 &  $-$18:19:47.45\\
Gaia Source ID & 94276941624384000 &  5702793425999272576 & 1049528378933767424 & 2393834386459511680\\
Parallax (mas) & $11.712 \pm 0.113$ &  $12.229 \pm 0.081$& $5.711 \pm 0.166$ & $10.582 \pm 0.153$\\
$\mu_{RA}$ (mas yr$^{-1}$) & $-213.07 \pm 0.14$  & $-133.22 \pm 0.07$ & $-115.14 \pm 0.13$ & $-0.58 \pm 0.14$\\
$\mu_{DEC}$ (mas yr$^{-1}$) & $-250.37 \pm 0.11$ & $+218.22 \pm 0.08$ & $-9.48 \pm 0.17$ & $-114.18 \pm 0.10$\\
$V_{\rm tan}$ (km s$^{-1}$) & $133.1 \pm 1.3$ & $99.1 \pm 0.7$ & $95.9 \pm 2.8$  &  $51.1 \pm 0.7$
\enddata
\tablecomments{J0205+2057 and J0958+5853 were not properly recognized in the literature.}
\end{deluxetable*}

To confirm their unusual nature, we obtained follow-up spectroscopy of J0831$-$2231 and J2340$-$1819 at the 6.5m MMT on UT 2023
Dec 9 and 11. We obtained $4\times5$ min back-to-back exposures of each target using the Blue Channel Spectrograph
\citep{schmidt89} with the 500 l mm$^{-1}$ grating and a $1.25\arcsec$ slit, providing wavelength coverage 3700 - 6850 \AA\
and a spectral resolution of 4.7 \AA. \citet{hollands20} detected low-level 0.4-0.6\% photometric variations in J0551+4135 at a single
period of 840 s. In order to search for spectroscopic variations at similar timescales (due to potential changes in the average surface
temperature or composition), we also obtained 1 min long back-to-back exposures of J0551+4135 over 30 minutes on UT 2023 Dec 8. 

\begin{figure}
\includegraphics[width=3.4in]{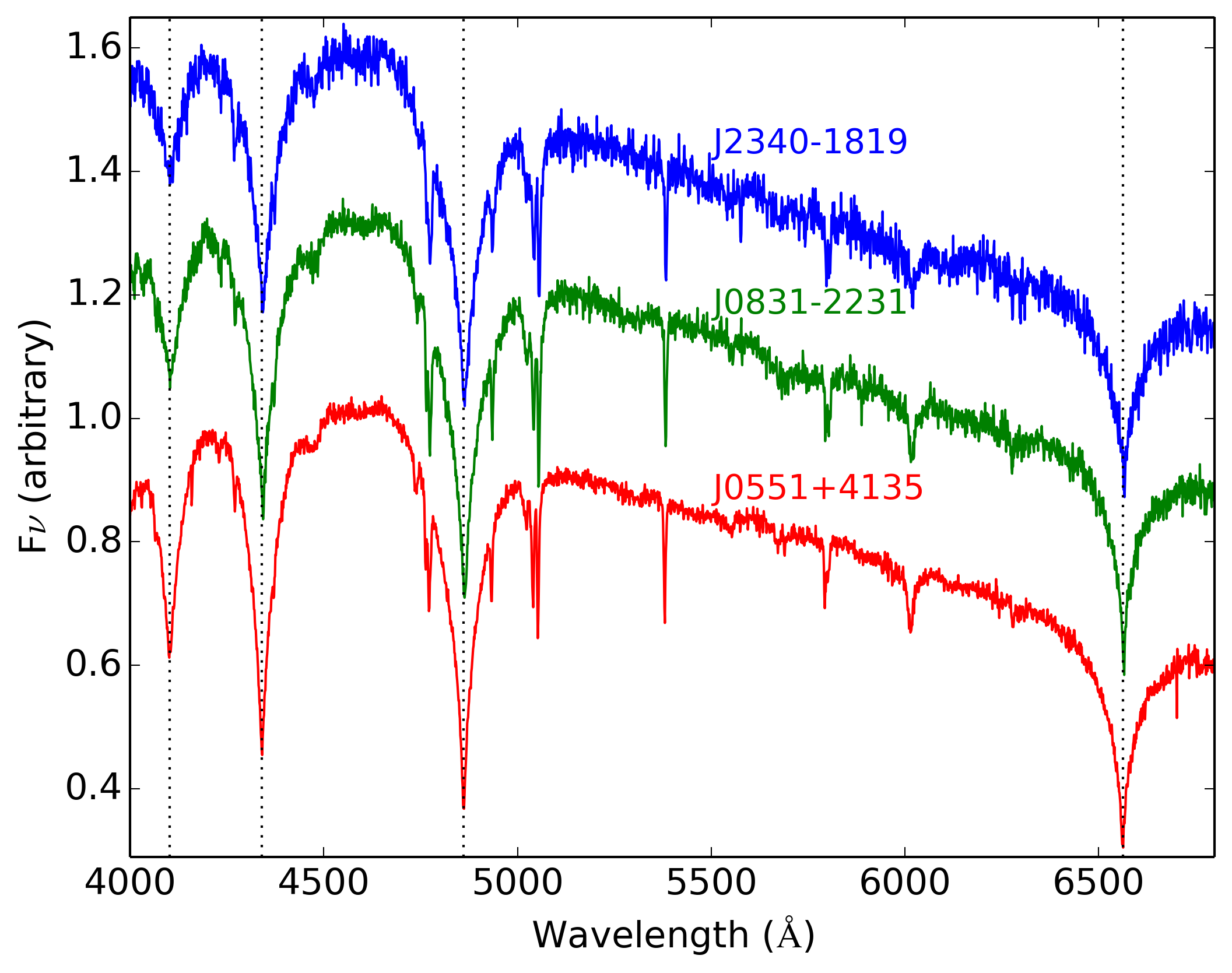}  
\caption{MMT spectra of the newly discovered DAQ white dwarfs J0831$-$2231 and J2340$-$1819 compared to the 
DAQ prototype J0551+4135. The spectra are normalized at 4500 \AA\ and arbitrarily shifted for display purposes. The dotted lines
mark the hydrogen Balmer series. All other features are from carbon.} 
\label{figmmt}
\end{figure}

Figure \ref{figmmt} shows the combined MMT spectra for J0831$-$2231 and J2340$-$1819 along with the DAQ prototype 
J0551+4135. The dotted lines mark the hydrogen lines; all other observed features are from carbon \citep{hollands20}.
The striking similarities between the spectra for these three stars confirm that J0831$-$2231 and J2340$-$1819 are also DAQ white dwarfs
with spectra dominated by hydrogen lines, and secondary features from carbon.

\section{Model Atmosphere Analysis}

\begin{figure}
\includegraphics[width=3.4in]{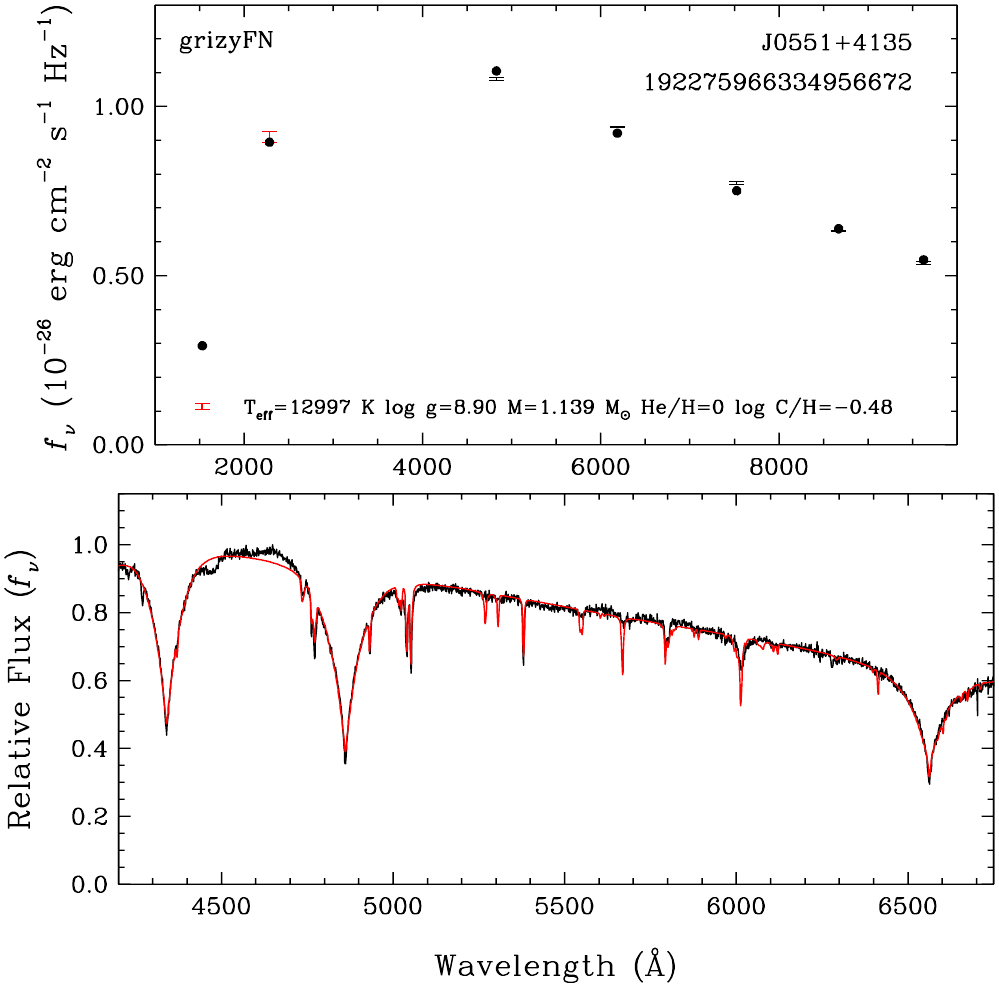}
\caption{Model atmosphere fits to the previously known DAQ white dwarf J0551+4135. The top and bottom panels show the
photometric and the spectroscopic fits, respectively. The best-fitting model parameters are presented in the top
panel, which also includes the Gaia DR3 Source ID, object name, and the photometry displayed in the panel.}
\label{fig0551}
\end{figure}

\begin{figure*}
\center
\includegraphics[width=3.4in]{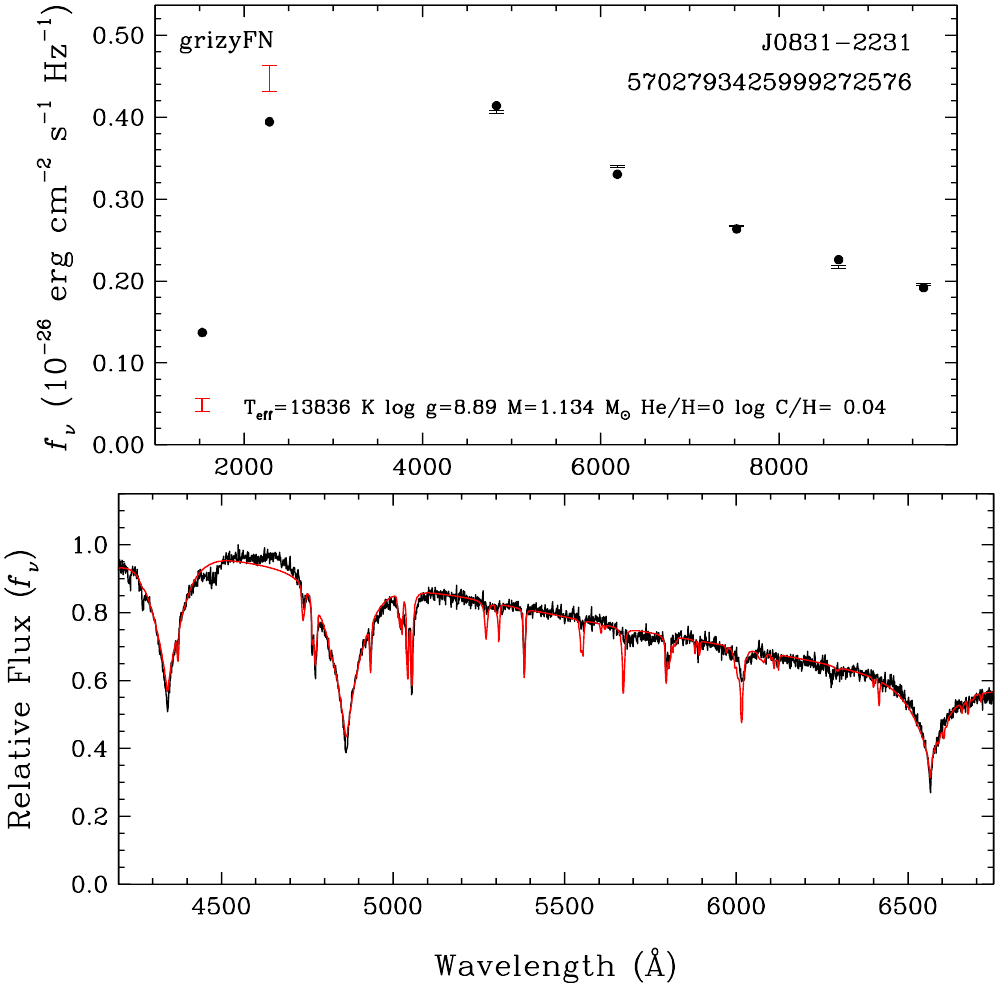}
\includegraphics[width=3.4in]{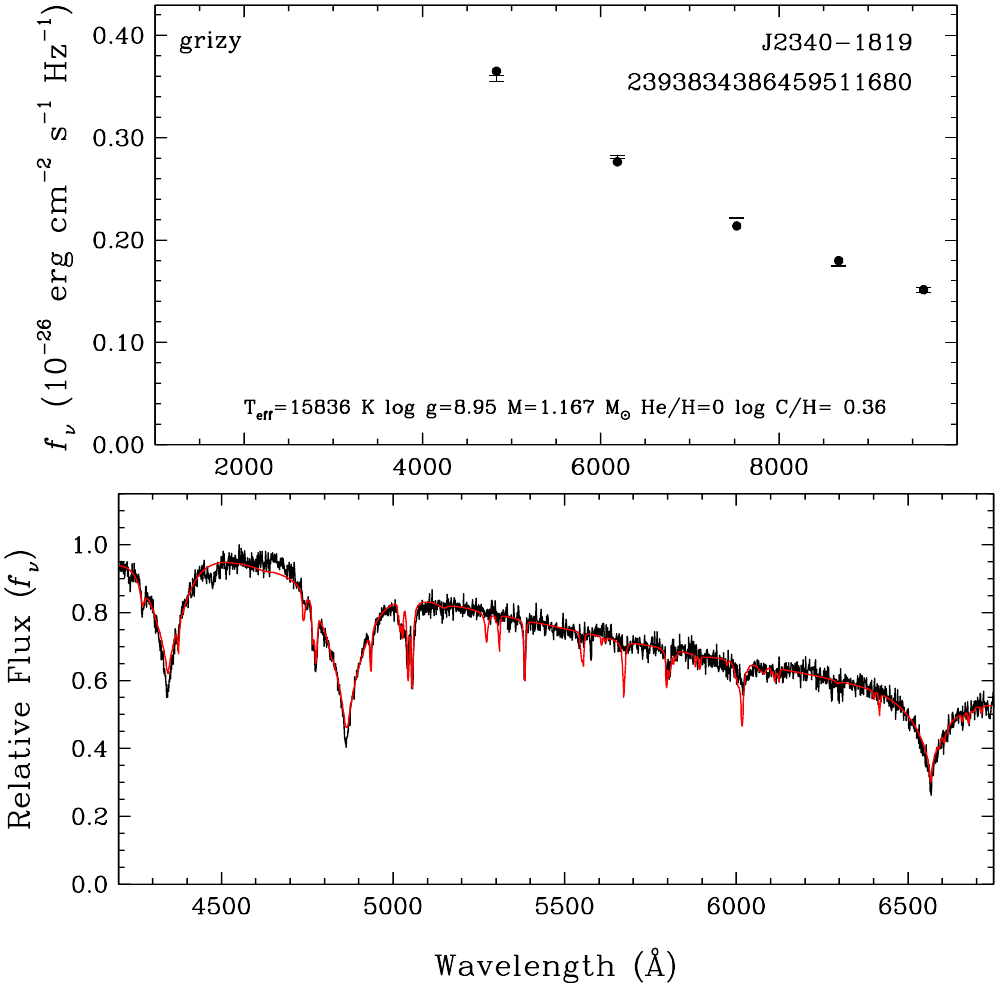}
\caption{Model atmosphere fits to two newly discovered DAQ white dwarfs. The symbols are the same as in Figure \ref{fig0551}.}
\label{figspec}
\end{figure*}

We rely on the photometric and spectroscopic techniques \citep{bergeron19} to constrain the physical parameters of our targets.
We use the SDSS $u$ (if available) and Pan-STARRS $grizy$ photometry along with the Gaia DR3 parallaxes to constrain
the effective temperature and the solid angle, $\pi (R/D)^2$, where $R$ is the radius of the star and $D$ is its distance. 
Since the distance is precisely known from Gaia parallaxes, we can constrain the radius of the star directly, and therefore the
mass based on the evolutionary models for white dwarfs. We use the geometric distances from \citet{bailer21} and the reddening values from
{\tt STILISM} \citep{capitanio17} for stars beyond 100 pc. 

We convert the observed magnitudes into average fluxes using the appropriate zero points, and compare with the average
synthetic fluxes calculated from model atmospheres with the appropriate chemical composition. A $\chi^2$ value is defined
in terms of the difference between the observed and model fluxes over all bandpasses,
properly weighted by the photometric uncertainties, which is then minimized using the nonlinear least-squares method of
Levenberg-Marquardt \citep{press86} to obtain the best fitting parameters. Given the abundances derived from the spectroscopic
fit, we repeat our photometric and spectroscopic fits until a consistent solution is found. The atomic data for carbon is relatively bad,
as the oscillator strengths for some of the lines are uncertain by 50\% or more. We exclude from our fits the carbon lines with
quality flags D and E in the NIST database, and also exclude two absorption features in the models near 5268 and 5668 \AA,
which are not observed.

For the purpose of this analysis, we rely on two distinct model atmosphere grids based on the calculations of \citet{blouin19}. The first one, more appropriate for the analysis of DAQ stars, covers the range $T_{\rm eff} = 12,000\ {\rm K}\ (500\ {\rm K})\ 17,000$ K, $\logg = 8.0\ (0.5)\ 9.5$, ${\rm He/H} = 0$, and $\log {\rm C/H} = -1.5\ (0.5)\ 2.5$ (where the numbers in parentheses indicate the step size). The second grid, more appropriate for the analysis of warm DQ/DQA stars, covers the range $T_{\rm eff} = 10,000\ {\rm K}\ (500\ {\rm K})\ 16,000$ K, $\logg = 7.0\ (0.5)\ 9.0$, $\log {\rm He/H} = 1.0\ (1.0)\ 4.0$, and $\log {\rm C/He} = -5.0\ (0.5)\ 1.0$. 

We rely on the evolutionary models described in \citet{bedard20} with CO cores, $q({\rm He})\equiv \log M_{\rm   He}/M_{\star}=10^{-2}$ and $q({\rm H})=10^{-10}$, which are representative of He-atmosphere (or thin H-atmosphere) white dwarfs. These evolutionary models are based on single star evolution, and the assumption of $q({\rm He})=10^{-2}$ is likely not representative for merger remnants. However, the helium buffer size has a relatively small effect on the mass-radius relation. For example, for a $T_{\rm eff}=15000$ K and $M=1.2~M_{\odot}$ star, $\logg$ changes by only 0.003 dex between the evolutionary models with $q({\rm He})=10^{-2}$ and $10^{-6}$. The helium buffer size also influences the cooling timescales of white dwarfs. However, we show below that our targets are likely delayed in their cooling by billions of years because of the $^{22}$Ne distillation \citep{blouin21,bedard24}. Hence, the cooling timescale differences due to a smaller helium buffer are negligible  compared to the multi-Gyr cooling delays due the distillation process.

\begin{deluxetable*}{cccccc}
\tablecolumns{6} \tablewidth{0pt}
\tablecaption{Physical Parameters for the DAQ white dwarfs. \label{tab2}}
\tablehead{\colhead{Parameter} & \colhead{J0205+2057} & \colhead{J0551+4135} & \colhead{J0831$-$2231} & J0958+5853 & \colhead{J2340$-$1819}}
\startdata
$T_{\rm eff}$ (K) & $16,427 \pm 228$ & $12,997 \pm 115$ & $13,836 \pm 180$ & $16,871 \pm 478$ & $15,836 \pm 291$ \\
$\logg$ & $9.01 \pm 0.02$ & $8.90 \pm 0.01$ & $8.89 \pm 0.01$ & $8.99 \pm 0.04$ & $8.95 \pm 0.02$ \\
Mass (\msun) & $1.194 \pm 0.008$ &  $1.139 \pm 0.005$ & $1.134 \pm 0.007$ & $1.184 \pm 0.019$ & $1.167 \pm 0.011$ \\
$\log$ C/H & $+0.97$ & $-0.48$ & $+0.04$ & $+0.91$ & $+0.36$ \\
Cooling age$^1$ (Gyr) & $0.96 \pm 0.04$ & $1.44 \pm 0.03$ & $1.23 \pm 0.04$ & $0.88 \pm 0.08$ & $0.97 \pm 0.05$
\enddata
\tablecomments{(1) Based on single star evolution and assuming CO cores. These cooling ages are most likely strongly underestimated compared to the true ages.}
\end{deluxetable*} 

Figure \ref{fig0551} shows our model fits to J0551+4135, the DAQ prototype. The top and bottom panels show our photometric and
spectroscopic fits, respectively. The top panel includes GALEX FUV and NUV photometry \citep[red error bars,][]{bianchi17} for
comparison. The best-fitting model has $T_{\rm eff}= 12,997 \pm 115$ K, $\logg = 8.90 \pm 0.01$,
$M = 1.139 \pm 0.005$ \msun, and $\log {\rm C/H} = -0.48$. Besides a small flux calibration issue near 4650 \AA,  this model
provides an excellent fit to the observed spectrum, including the Balmer lines and the strongest carbon features. \citet{koester19}
and \citet{hollands20} discuss in detail the issues with the accuracy of the oscillator strengths for the carbon lines, especially
below 4500 \AA. Our best-fitting model under-predicts the carbon line depths in the blue. However, our best-fitting model parameters
are consistent with the analysis presented in \citet{hollands20}, who found $T_{\rm eff}= 13,370 \pm 330$ K, $\logg = 8.91 \pm 0.01$,
and $\log {\rm C/H} = -0.83$. 

A striking feature in J0551+4135's observed spectrum is the flux depression near 4470 \AA. Normally we would associate that feature with the
neutral helium line at 4471 \AA, which is commonly observed in DB white dwarfs. However, if that feature is really due to \ion{He}{1}, then
we should also see a stronger \ion{He}{1} absorption feature at 5876 \AA, which is clearly absent. We explored model atmospheres
with varying helium abundances, and found that it is impossible to have a strong 4471 \AA\ feature and hide the one at 5876 \AA.
\citet{hollands20} put an upper limit of $\log {\rm He/H}<-0.3$ based on the absence of the latter. They also associate the feature
around 4470 \AA\ with carbon, though there was a typo in their paper and this feature was erroneously reported as 4570 \AA\
(M. Hollands 2024, private communication). Unfortunately, there are significant uncertainties in the oscillator strengths for the carbon lines
in the blue; for example, \citet{koester19} find factor of $\sim$4 differences between the NIST, VALD, and the literature values for the
\ion{C}{1} $\lambda$4270.221 \AA\ line. Hence, we are not able to resolve the issue with the 4470 \AA\ feature at this point, but it is clearly not from
\ion{He}{1}, and is most likely from the carbon triplet at 4467.714, 4478.727, and 4479.840 \AA\ (vacuum). Regardless of these issues
with matching the carbon lines below about 4500 \AA, we confirm that J0551+4135 has a hydrogen dominated atmosphere with
significant amounts of carbon present, and that there is no evidence of any helium in its spectrum.

Figure \ref{figspec} shows our model fits to the newly discovered DAQ white dwarfs J0831$-$2231 and J2340$-$1819.  Even though the
problems in matching some of the carbon lines persist, like in J0551+4135, carbon and hydrogen atmosphere models provide an excellent
match to the overall spectra of both targets. Table \ref{tab2} presents the best-fitting model parameters for each source. These two
stars are slightly hotter and even more carbon-rich than J0551+4135, but otherwise they have similar masses and estimated
cooling ages of $\sim$1 Gyr. As discussed above, these ages are based on standard single-star evolution, and are likely strongly underestimated.

\section{Two Additional DAQ White Dwarfs Hiding in the Literature}

\begin{figure}
\center
\includegraphics[width=3.4in]{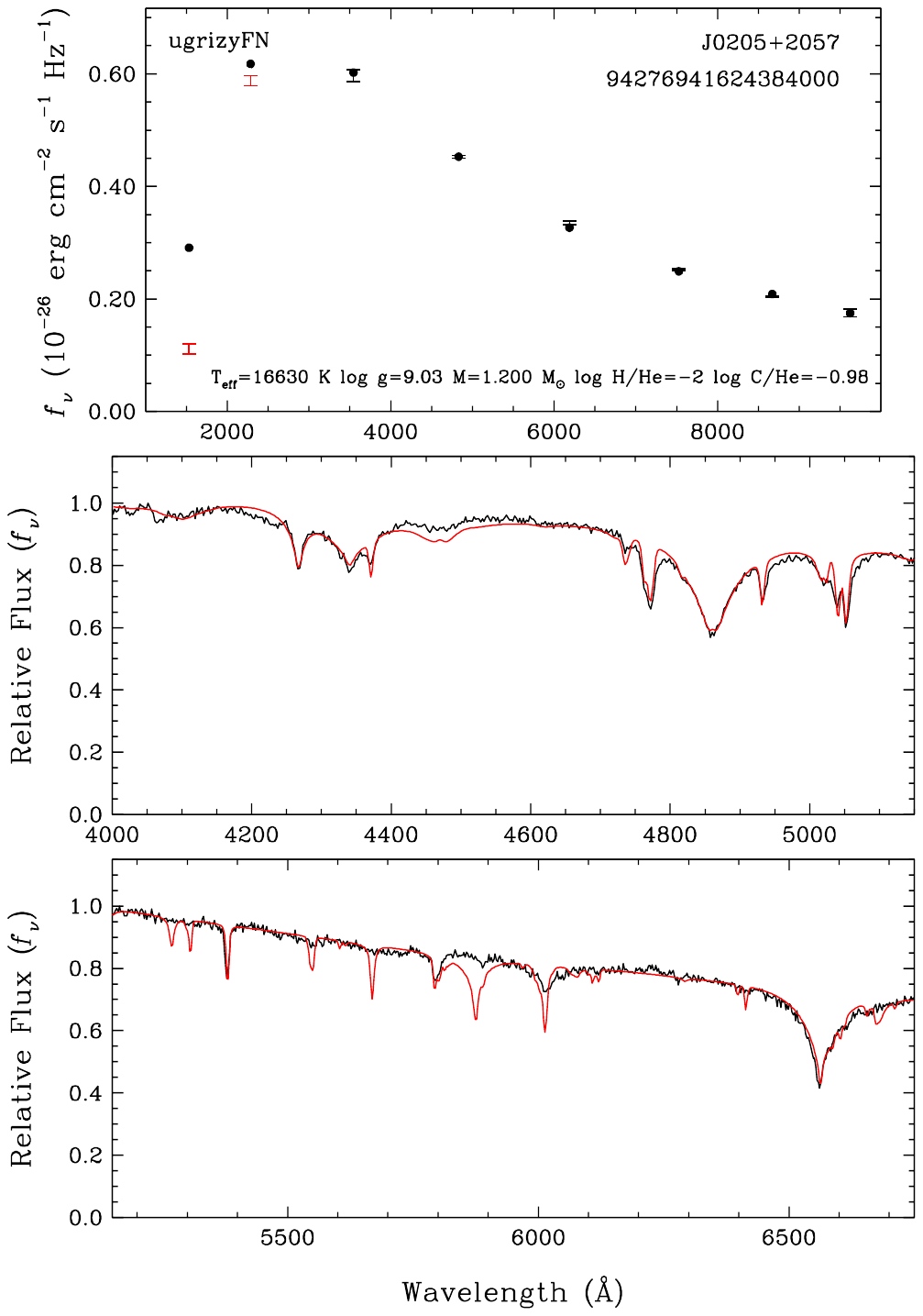}
\caption{Model atmosphere fits to J0205+2057 under the assumption of a helium-dominated atmosphere. Note the relatively strong \ion{He}{1} $\lambda$5876 \AA\ feature predicted in the models that is not observed in the spectrum of this object presented in the bottom panel.
This indicates that J0205+2057 does not have a helium-dominated atmosphere. In fact, there is no evidence of helium in the spectrum
of this DAQ white dwarf.}
\label{fighe}
\end{figure}

\begin{figure*}
\includegraphics[width=3.4in]{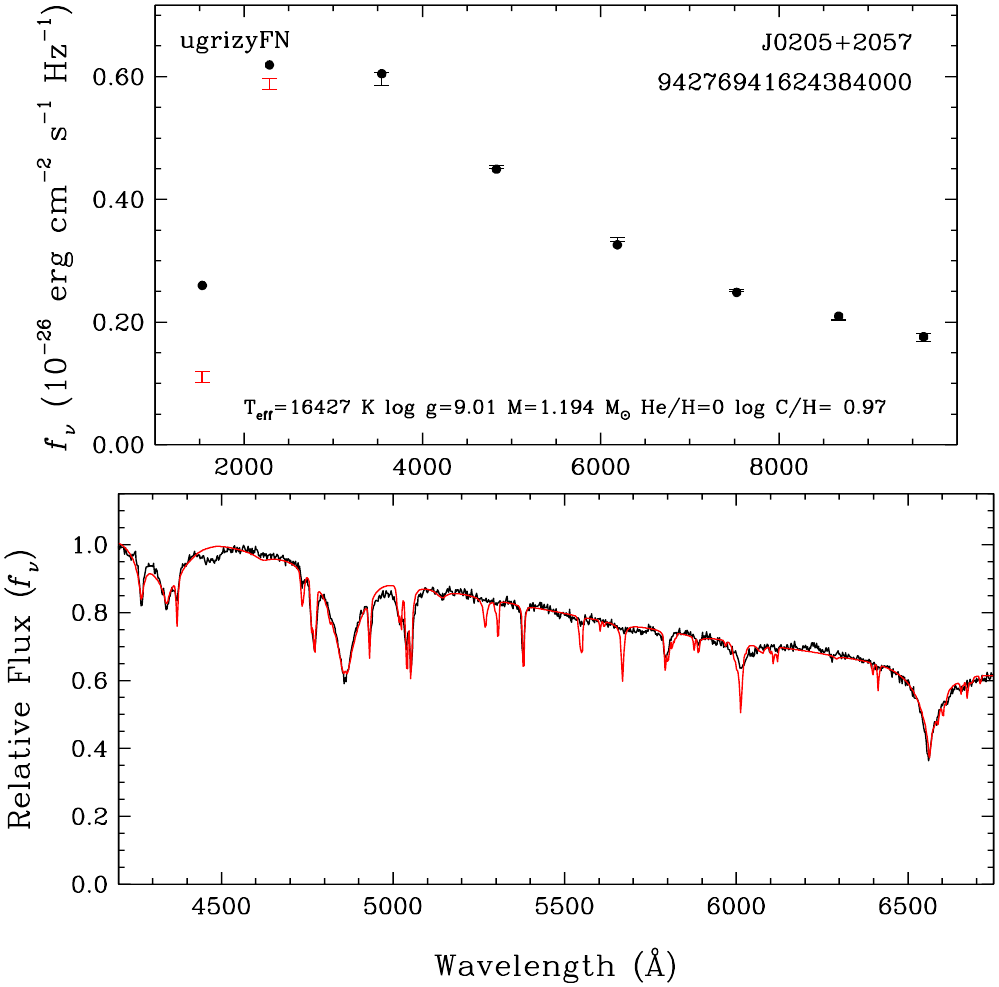}
\includegraphics[width=3.4in]{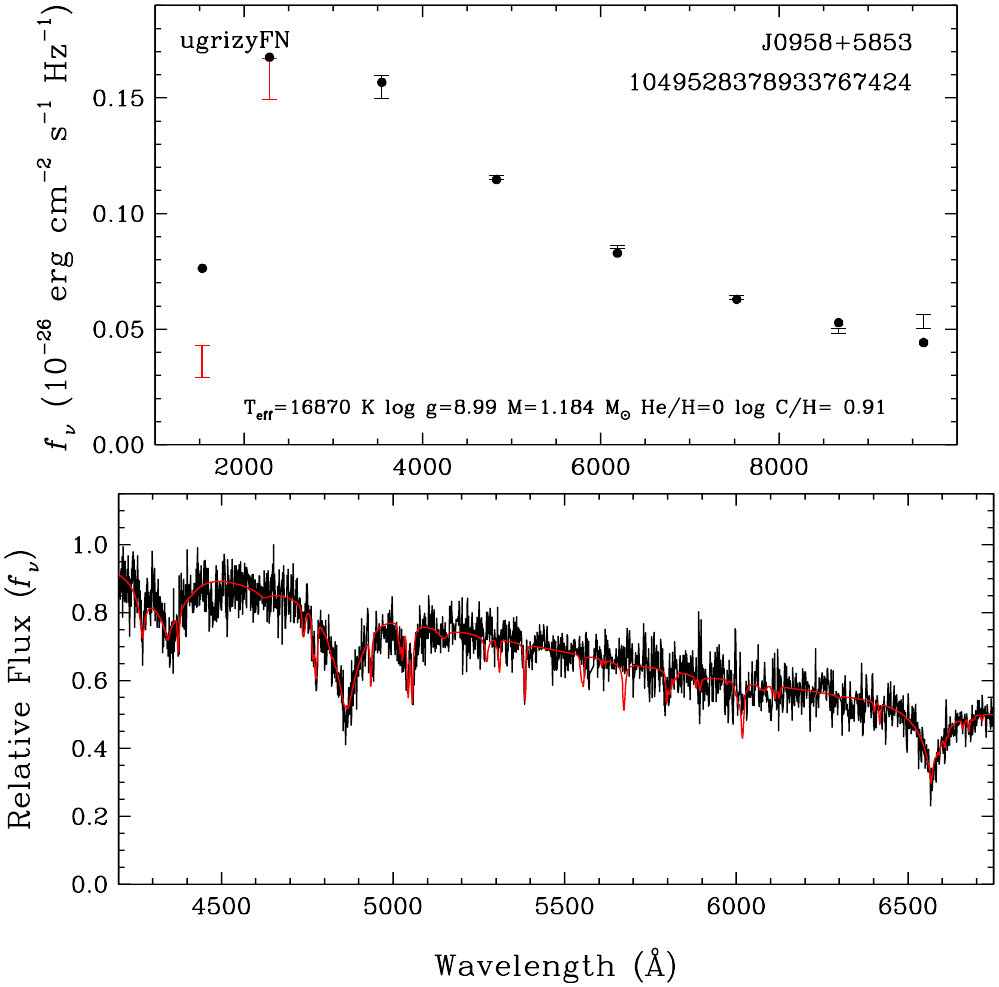}
\caption{Mixed carbon and hydrogen atmosphere model fits to two DAQ white dwarfs that were not properly recognized in the literature.}
\label{figpast}
\end{figure*}

The discovery of two new DAQs in our follow-up sample (Jewett et al., in prep.) prompted us to revisit the spectral
classification of warm DQ white dwarfs. Through this process, we identified two additional DAQ white
dwarfs that were not properly recognized in the literature: J0205+2057 (G35-26) and J0958+5853 (SDSS J095837.00+585303.0).

\citet{liebert83} identified J0205+2057 as the first white dwarf to show atomic lines of both hydrogen and carbon, and classified it as a DAQ3 white dwarf in the spectral classification system of \citet{sion83}. He noted that it is tempting to identify this star with the hot end of the helium-rich white dwarfs that show traces of dredged-up carbon (DQ type), but his search for  the \ion{He}{1} $\lambda$5876 \AA\ line was unsuccessful. Based on a model atmosphere analysis, \citet{thejll90} concluded that J0205+2057 has a helium-dominated atmosphere with a temperature between 11,000 and 14,000 K and a hydrogen abundance of 0.5 to 1\% by number. However, this analysis was based on a blue spectrum that did not include the \ion{He}{1} $\lambda$5876 \AA\ line, and they erroneously associated the broad feature at 4470 \AA\ with \ion{He}{1}. 

Figure \ref{fighe} shows the optical photometry and spectroscopy of J0205+2057 along with our model atmosphere fits
under the assumption of a helium dominated atmosphere ($\log {\rm H/He}=-2$). First of all, the spectral classification of white dwarfs solely depends
on the observed spectrum \citep{sion83}, and not the composition determined from a model atmosphere analysis. Here the
Balmer lines are the strongest features, and the carbon lines are secondary. Hence, J0205+2057 is clearly a DAQ white dwarf, and as
\citet{liebert83} noted, it is the first DAQ white dwarf ever found. Second, given the severe problems with the atomic data for carbon
\citep[see e.g.][]{koester19}, the broad feature from carbon lines near 4470 \AA\ is easily confused with the \ion{He}{1} feature at 4471 \AA,
leading to incorrect atmospheric composition measurements. Our spectroscopic model fit under the assumption of a helium-dominated
atmosphere provides a decent fit to the observed 4470 \AA\ feature, but it predicts an even stronger helium feature at 5876 \AA\ that
is clearly not observed. The \ion{He}{1} $\lambda$5876 \AA\ feature is the strongest feature in DB white dwarfs, and it is the last helium feature to disappear (see for example the infamous case of GD 362 in \citealt{zuckerman07}). 

Figure \ref{figpast} shows the carbon and hydrogen atmosphere model fits for J0205+2057 and another DAQ overlooked in the literature, J0958+5853. Both of these stars have Balmer lines that are stronger than the carbon lines, hence are clearly DAQ white dwarfs.
Here the DAQ models provide an excellent match to the spectra of both targets, and all of the arguments presented here for J0205+2057 also apply to the spectrum of J0958+5853. Besides the obvious problems with matching some of the carbon lines, as in the other DAQs discussed above, the Balmer lines and the carbon
features are fit fairly well by models with $\log {\rm C/H} \approx0.9$. Adding J0205+2057 and J0958+5853 to the list, we now have a class of DAQ white
dwarfs with at least five members, four of which are in the 100 pc sample.

\section{The Distinction Between DAQ and DQA White Dwarfs}

\begin{figure*}
\includegraphics[width=3.4in]{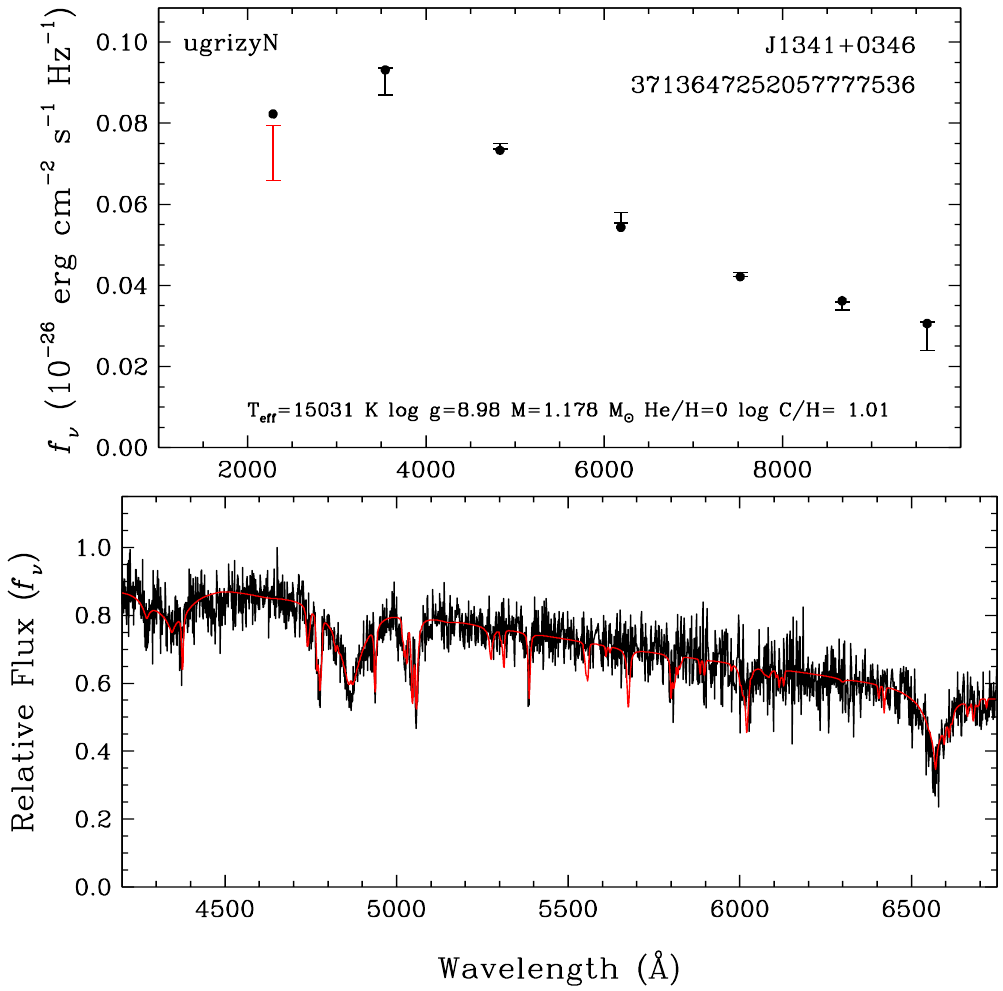}
\includegraphics[width=3.4in]{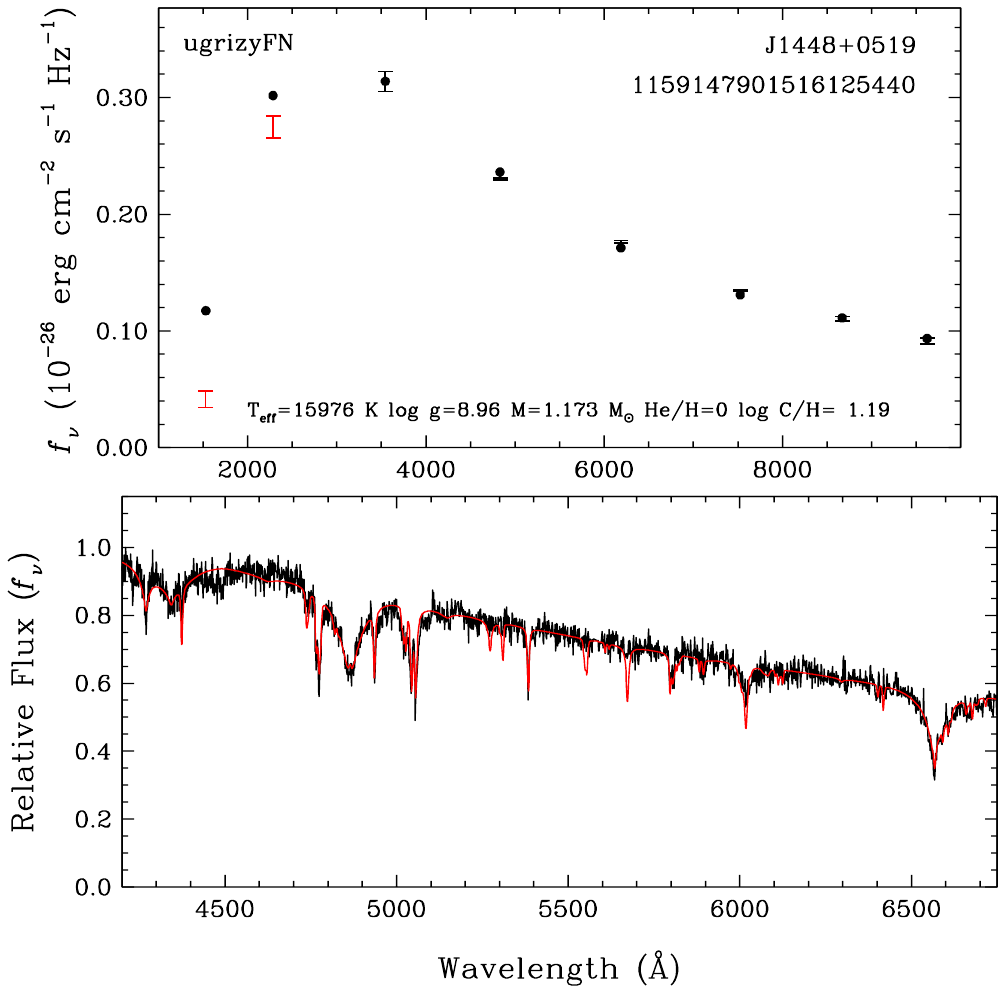}
\caption{Examples of carbon and hydrogen atmosphere model fits to two warm DQA white dwarfs with strong Balmer lines. These objects are
J1341+0346 (SDSS J134124.28+034628.7) and J1448+0519 (WD 1446+055).}
\label{figdqa}
\end{figure*}

Many of the warm DQ white dwarfs show evidence of hydrogen in their spectra \citep{koester19,coutu19}, and several of them
display relatively strong H$\alpha$ and H$\beta$ lines.  Figure \ref{figdqa} shows the spectral energy distributions and our model
fits for two such objects. The spectra for these two stars appear relatively similar to the DAQ stars discussed above. However,
we found it difficult to classify these objects; the carbon lines are slightly deeper than the H$\beta$ line, but it is hard
to tell if they are also deeper than the H$\alpha$ line. These stars were classified either as DQ or DQA in the literature \citep{kepler15,coutu19}.

Traditionally, warm DQ and DQA white dwarfs are expected to have helium-dominated atmospheres. However, we now know that the broad feature
at 4470 \AA\ is not from helium and it is most likely due to carbon, as none of these stars display the much stronger \ion{He}{1} $\lambda$5876 line.
Analyzing the spectral energy distributions of warm DQs with parallax measurements under the assumption of helium-dominated
atmospheres, \citet{koester19} found that they can constrain the C/He ratio in these atmospheres relatively well as long as
$\log {\rm C/He}\leq -1$. However, at higher ratios the errors become as large as 0.7 dex. In addition, in two cases, including J1448+0519
shown here, their best-fit models with helium-dominated models predict strong helium lines that are not observed (see their Figure 1).
They could not constrain the abundances for those two stars, but they were able to put an upper limit on the He/C ratio from the
absence of \ion{He}{1} $\lambda$5876. However, given the lack of any helium lines in the spectra of these stars, it is also
possible to fit warm DQA stars with no helium, as we did for the DAQ stars. 

\begin{figure}
\center
\includegraphics[width=3.4in]{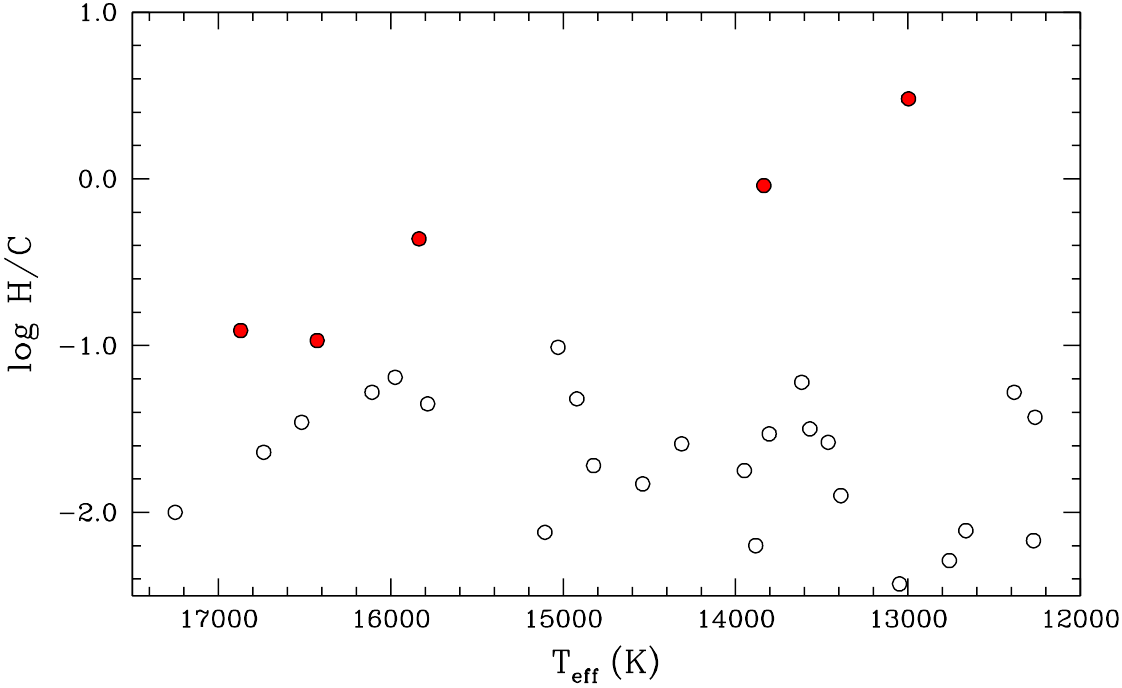}
\caption{H/C ratio versus effective temperature for warm DQ white dwarfs. Red points mark the 5 DAQ white dwarfs identified here.}
\label{figratio}
\end{figure}

Figure \ref{figdqa} shows the results from this experiment for these two DQA white dwarfs. Carbon and hydrogen atmosphere models
provide fits that are just as good as the fits using the carbon and helium (and trace amounts of hydrogen) atmospheres. The spectral energy distributions of these two stars
can be explained by models with $T_{\rm eff}=$ 15,000-16,000 K and $\log$ C/H ranging from +1.01 to +1.19. These parameters
are similar to the temperatures derived for the DAQ white dwarfs discussed above, and the C/H ratios are slightly higher. 
Note that we are not claiming that these DQA white dwarfs have no helium, we simply cannot tell if there is any.
Due to the absence of helium features in their spectra, we can at best put an upper limit on the He/C ratio, but this ratio could be as low as
zero. 

Figure \ref{figratio} shows the H/C ratio for the warm DQ white dwarfs from \citet{koester19} and \citet{coutu19}, as well as the 100 pc sample of
Jewett et al. (in prep.), under the assumption of carbon and hydrogen atmospheres with no helium. We restrict this plot to the temperature
and abundance ranges covered by our DAQ model grid. Note that the assumption of a helium-free atmosphere has minimal impact on the derived H/C ratios for these stars since no helium lines are observed. For reference, the helium-dominated and helium-free model fits shown in
Figures \ref{fighe} and \ref{figpast} for J0205+2057 have $\log {\rm H/C}=-1.02$ and $-0.97$, respectively.

Figure \ref{figratio} shows that DAQ white dwarfs are the most hydrogen-rich stars among the warm DQ sample, but otherwise
they belong to the same population. Regardless of whether there is helium or not in the atmosphere, this figure demonstrates that
the distinction between DAQ and DQA white dwarfs is superficial; warm DQ white dwarfs simply display a range of hydrogen abundances
in their atmospheres, and depending on how much hydrogen is present and which lines are stronger, we classify them as either a DAQ, DQA, or DQ. 

\section{Discussion}

\subsection{Kinematics}

\citet{hollands20} noted the unusual kinematics of J0551+4135 for its cooling age. J0551+4135 has a tangential velocity of only 30 km
s$^{-1}$. However, including its gravitational redshift corrected radial velocity of $-114$ km s$^{-1}$, its total velocity with
respect to the local standard of rest is $129 \pm 5$  km s$^{-1}$. The remaining four DAQs have larger tangential
velocities (see Table \ref{tab1}): all have $V_{\rm tan} \geq 51$ km s$^{-1}$, and three have $V_{\rm tan} \geq96$ km s$^{-1}$, pointing to a kinematically old thick disk or halo population.

We measure radial velocities of $-2.8 \pm 12.6$, $142.1 \pm 18.9$, and $128.8 \pm 20.0$ km s$^{-1}$ from our MMT spectra for
J0551+4135, J0831$-$2231, and J2340$-$1819, respectively. The former is consistent with the radial velocity measurement (before
the gravitational redshift correction) of $+2.7 \pm 5.1$ km s$^{-1}$ from \citet{hollands20}. We measure radial velocities of
$-18.3 \pm 41.9$ and $203.2 \pm 21.4$ km s$^{-1}$ for J0205+2057 and J0958+5853, respectively, using their SDSS spectra. Given the relatively
large masses of these white dwarfs and the gravitational redshift corrections of $\approx120$ km s$^{-1}$, J0205+2057,
J0551+4135, and J0958+5853 have a relatively large line-of-sight motion. 

\begin{deluxetable}{cccc}
\tablecolumns{4} \tablewidth{0pt}
\tablecaption{ $UVW$ velocities for the DAQ sample. \label{tab3}}
\tablehead{\colhead{Object} & \colhead{$U$} & \colhead{$V$} & \colhead{$W$} \\
 & (km s$^{-1}$) & (km s$^{-1}$) & (km s$^{-1}$) }
\startdata
J0205+2057   & $198  \pm 27$ & $-64 \pm 19$& $4 \pm 26$ \\ 
J0551+4135   & $130 \pm 12$  & $-8 \pm 2$  & $21 \pm 2$  \\
J0831$-$2231 & $-89 \pm 8$   & $31 \pm 17$ & $18 \pm 3$  \\
J0958+5853   & $-109 \pm 13$ & $14 \pm 7$  & $4 \pm	16$ \\
J2340$-$1819 & $35 \pm 3$    & $-31 \pm 5$ & $-5 \pm 19$  
\enddata
\end{deluxetable}

\begin{figure}
\includegraphics[width=3.4in]{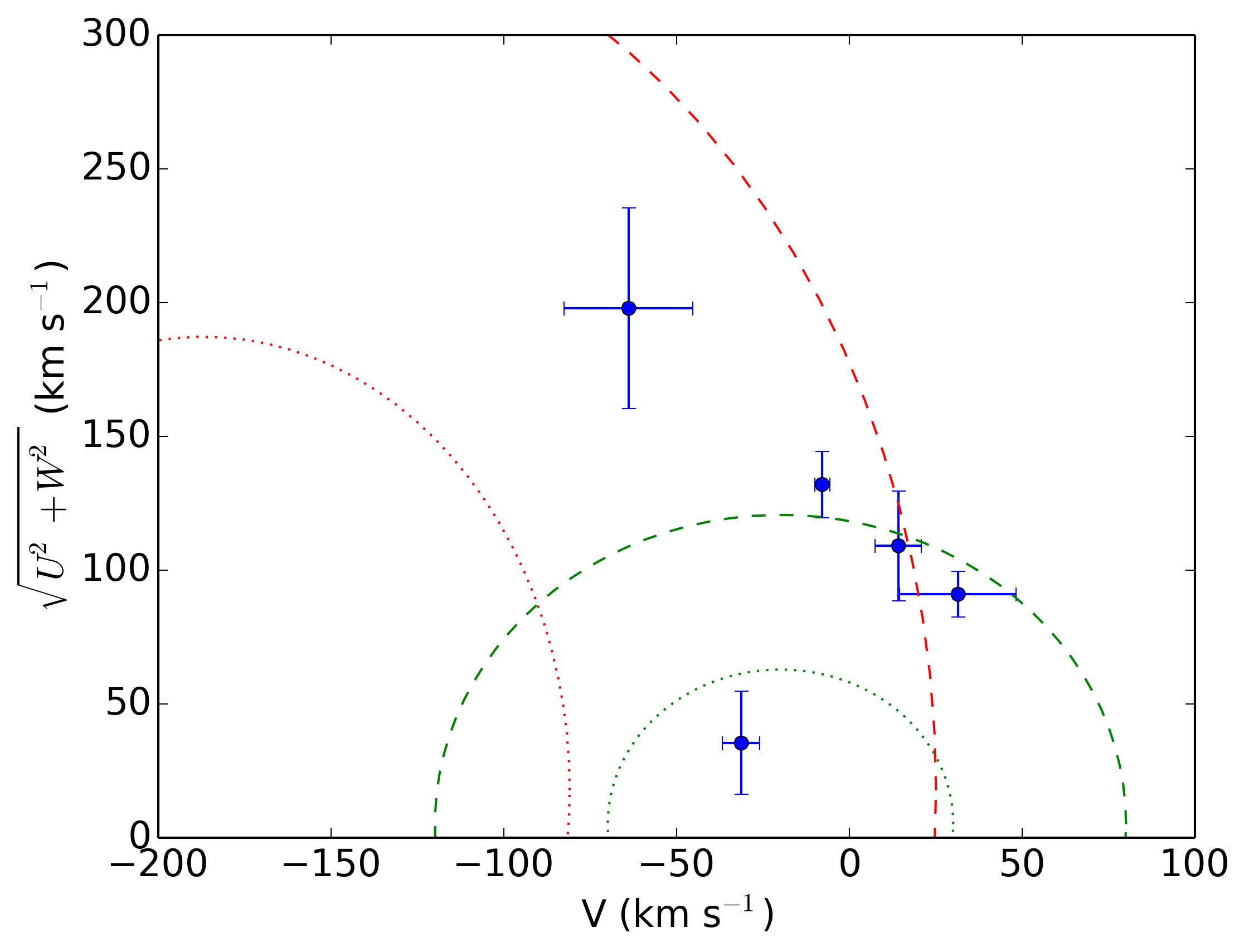}
\caption{Toomre diagram for the DAQ white dwarf sample. The dotted and dashed lines show the $1\sigma$ and $2\sigma$
velocity ellipsoids for the thick-disk (green) and halo (red lines), respectively \citep{chiba00}.}
\label{uvw}
\end{figure}

We compute the $UVW$ velocities using the Gaia parallax and zero point correction, and correct the values to the local standard of rest \citep{schonrich10}.
Table \ref{tab3} presents these velocities. Four of the DAQs orbit with the Sun ($|V|\leq31$ km s$^{-1}$) and all five have small vertical
velocities ($|W|\leq$ 21 km s$^{-1}$). However, the $U$ components of motion (toward/away from the Galactic Center) for these stars are relatively large.
J0205+2057 is moving towards the Galactic Center at $198  \pm 27$ km s$^{-1}$ and J0551+4135 at $130 \pm 12$ km s$^{-1}$, whereas J0958+5853
is going in the opposite direction at $-109 \pm 13$ km s$^{-1}$.
Figure \ref{uvw} plots the distribution of Galactic U, V, and W velocity components for our five targets, along with the
$1\sigma$ (dotted) and $2\sigma$ (dashed) velocity ellipsoids for the thick-disk and halo \citep{chiba00}. 
J0205+2057 is likely a halo member, whereas J0831$-$2231, J0958+5853, and J2340$-$1819 are likely thick disk objects. J0551+4135 is more ambiguous,
but its small V and W velocities suggest it is probably a thick disk object as well; it is only 46 pc away.
Clearly these five DAQs belong to a kinematically old population, even though their estimated cooling ages are of order 1 Gyr
under the assumption of single star evolution.

\subsection{Rapid Rotation in J0831$-$2231 and J2340$-$1819}

\begin{figure*}
\center
\includegraphics[width=3.4in]{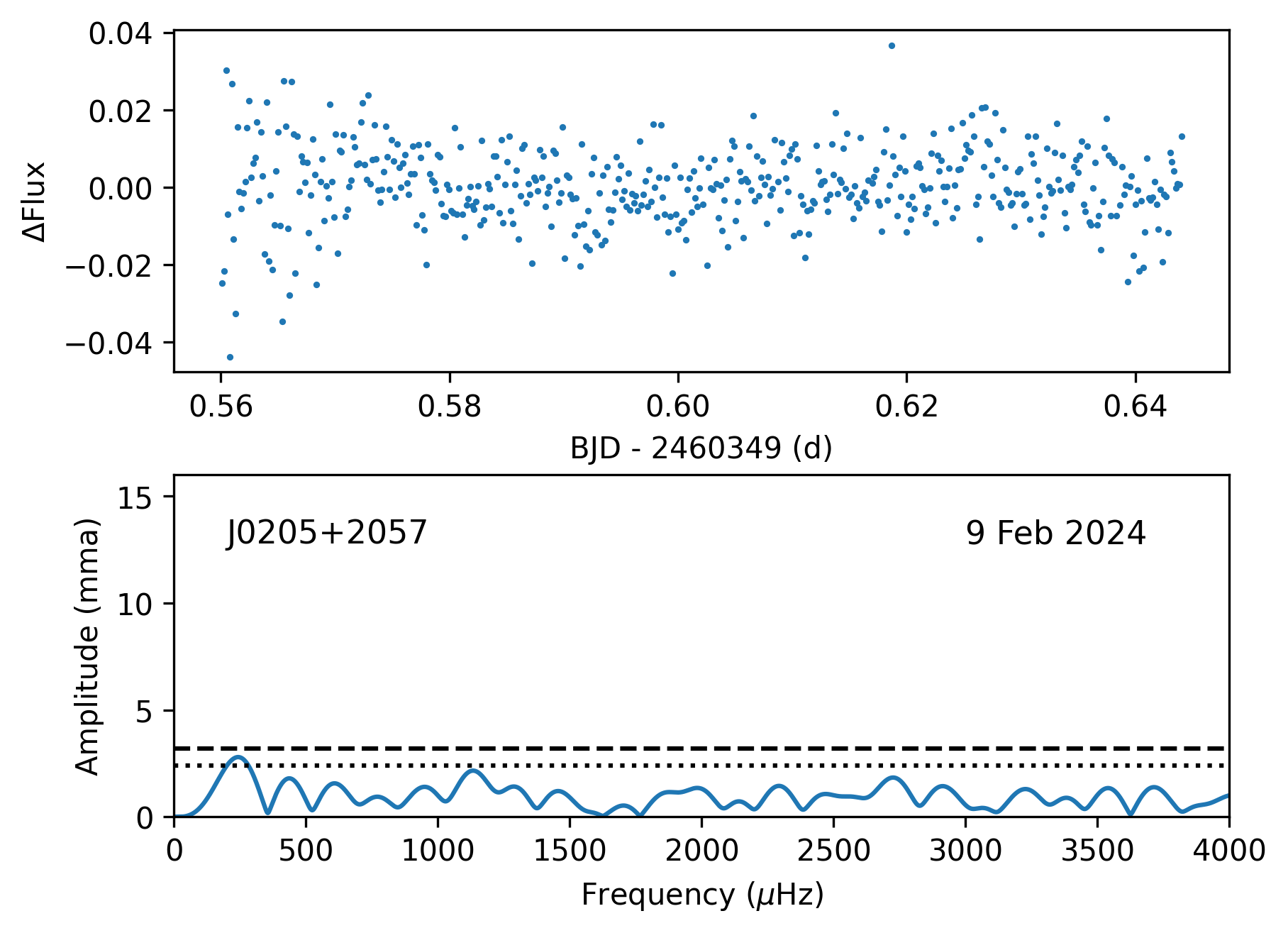}
\includegraphics[width=3.4in]{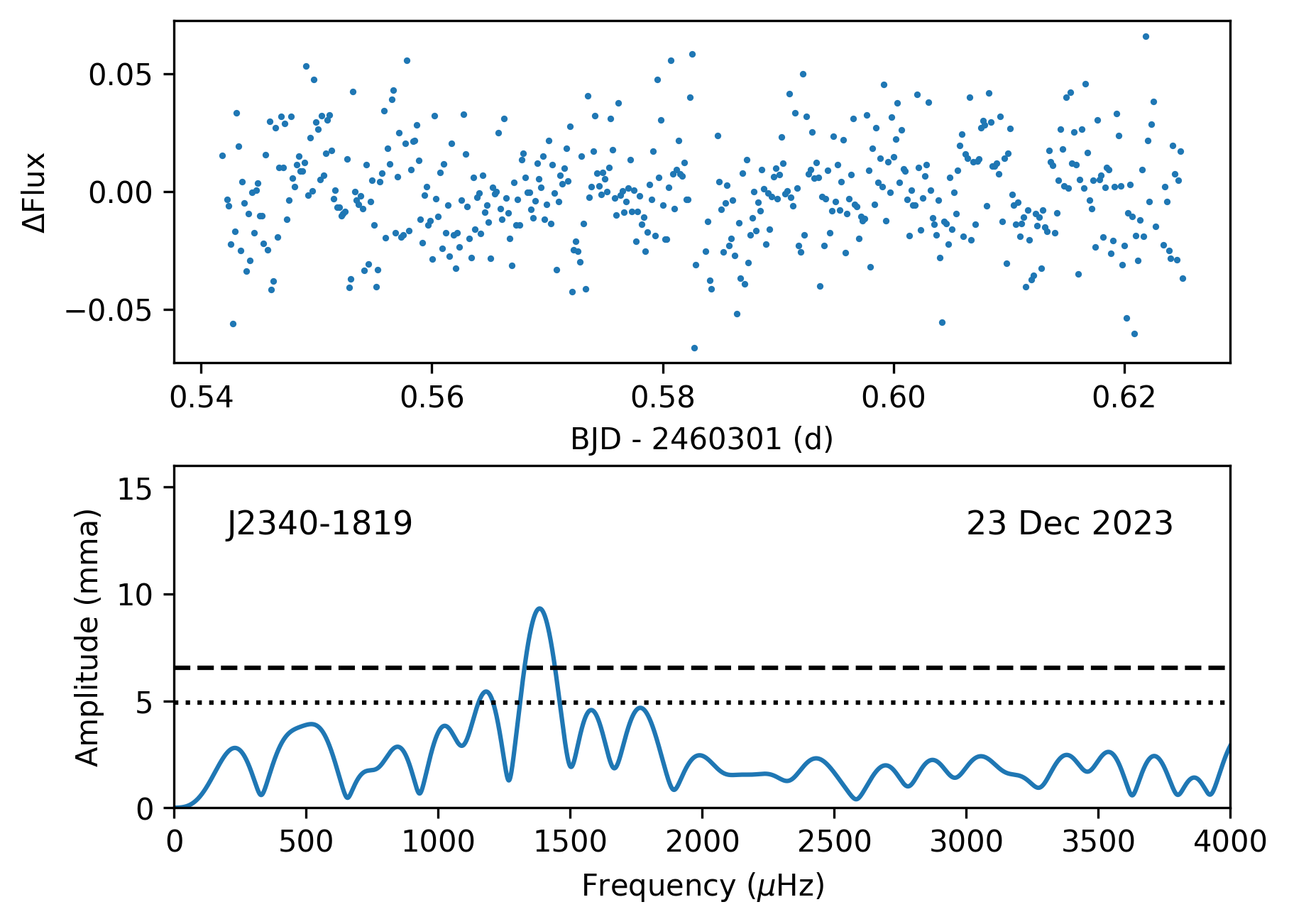}
\includegraphics[width=3.4in]{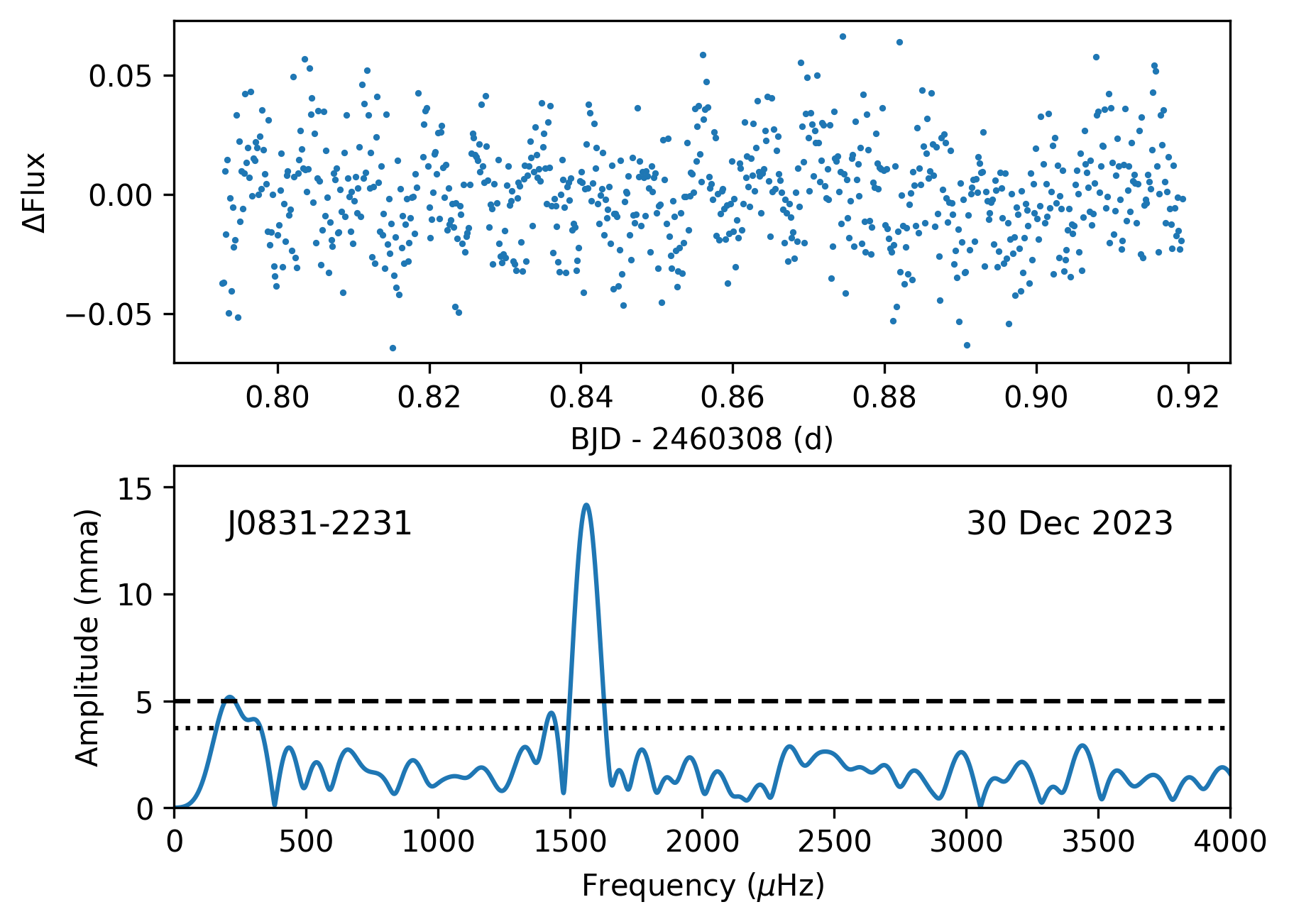}
\includegraphics[width=3.4in]{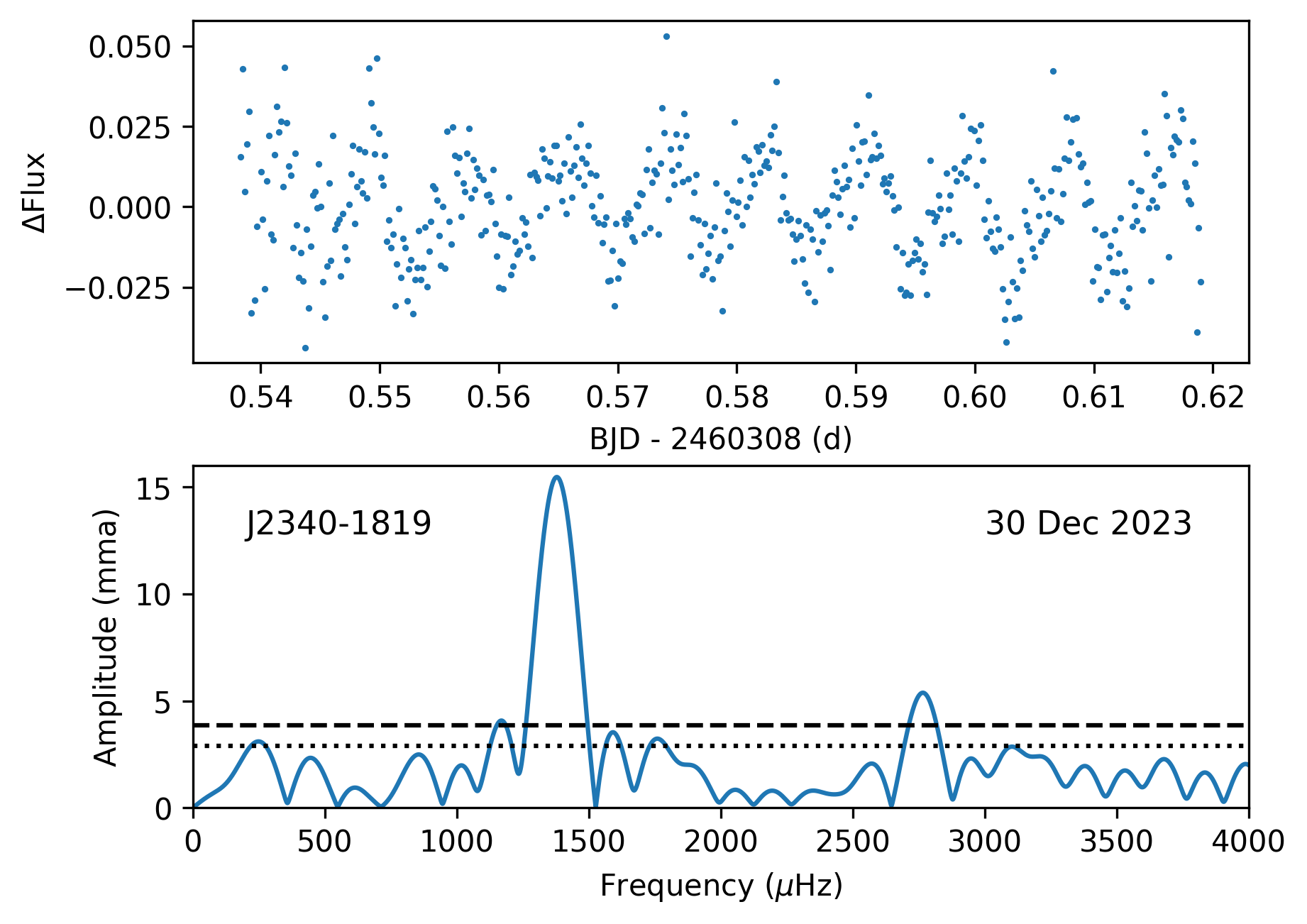}
\caption{APO time-series photometry of three DAQ white dwarfs (top panels) and their Fourier transforms (bottom panels).
The dotted and dashed lines show the 3$\langle {\rm A}\rangle$ and  4$\langle {\rm A}\rangle$ levels. J2340$-$1819 was
observed on two different nights (right panels).}
\label{photapo}
\end{figure*}

We acquired high speed photometry of three of the DAQs, J0205+2057, J0831$-$2231 and J2340$-$1819, on UT 2023
Dec 23, Dec 30, and 2024 Feb 9 using the APO 3.5m telescope with the Astrophysical Research Consortium Telescope
Imaging Camera (ARCTIC) and the BG40 filter. We obtained back-to-back exposures of 10 s over $\geq2$ hours for each
target. We binned the CCD by $3\times3$, which resulted in a plate scale of $0.34\arcsec$ pixel$^{-1}$ over a field of view
of 7.85 square arc minutes. This setup results in an overhead of 4.5 s for each exposure, resulting in a cadence of 14.5 s. 

Figure \ref{photapo} shows the APO light curves and their Fourier transforms. There is no evidence of short term variability
in J0205+2057's (top left) light curve at the 4$\langle {\rm A}\rangle$ level, where $\langle {\rm A}\rangle$ is the average amplitude
in the Fourier transform. However, the other two stars presented in this figure are clearly variable.
The bottom left panels show the lightcurve of J0831$-$2231 obtained over 3 hours on 30 Dec 2023, along with
its Fourier transform. There is only a single significant peak detected at a frequency of $1560.6 \pm 3.5~\mu$Hz with
$14.1 \pm 1.0$ milli-modulation amplitude (mma). J0831$-$2231 is outside of the ZZ Ceti instability strip. In addition, since
only a single mode is detected, the variability must be due to rotation along with either a spotted surface or an
inhomogeneous atmosphere. Regardless of the exact cause of variability, J0831$-$2231 is rotating with a period of only 10.7 min.

The right panels in Figure \ref{photapo} show the APO light curves of J2340$-$1819 from two different nights separated by
a week. The light curves cover 120 and 116 min, respectively. There is only a single significant peak detected on 23 Dec 2023
at a frequency of $1383.8 \pm 10.6~\mu$Hz with an amplitude of $9.3 \pm 1.3$ mma. The data from 30 Dec 2023 also shows a single
significant frequency at $1378.9 \pm 3.8~\mu$Hz with an amplitude of $15.7 \pm 0.7$ mma, and its first harmonic at a
frequency of $2764.7 \pm 11.0~\mu$Hz. J2340$-$1819 is clearly outside of the ZZ Ceti instability strip, and rotates rapidly with
a period of 12 min. 

J0551+4135 is the only DAQ within the boundaries of the ZZ Ceti instability strip \citep[see Figure 4 in][]{kilic23b}. Carbon does
not affect the excitation of $g$ modes in DAQ white dwarfs since the excitation is due to the $\kappa$ mechanism by the dominant
chemical species, in this case, hydrogen (A. C\'orsico 2024, private communication). Both \citet{vincent20} and \citet{hollands20}
reported the discovery of pulsations in J0551+4135 with a single frequency peak. The latter obtained time-series photometry
over several nights, and found that this single peak changed both in frequency (from 1186 to 1202 $\mu$Hz) and amplitude
(by a factor of two) over a 10 day period. This change in frequency is possible for pulsations, but not for rotation. \citet{hollands20}
suggest that the change in amplitude could be due to unresolved rotational splitting. 

Our time-series spectroscopy obtained over 30 min for J0551+4135 does not provide any additional constraints on the rotation period.
Fitting each 1 min long spectroscopic exposure, we find that the $\log$ (C/H) ratio of the best-fitting model to the spectra ranges from
$-0.54$ to $-0.44$ with an average of $-0.48 \pm 0.02$ dex. We obtained additional time-series photometry of J0551+4135 at the
APO 3.5m telescope and detected multi-mode pulsations over a few nights. However, the number of detected modes is still relatively
small, and it does not enable us to perform detailed asteroseismology of this object yet. Hence, the rotation periods of J0551+4135
and J0958+5853 remain currently unconstrained. 

\subsection{The Origin of the DAQ Class}

\begin{figure*}
\center
\includegraphics[width=6.4in]{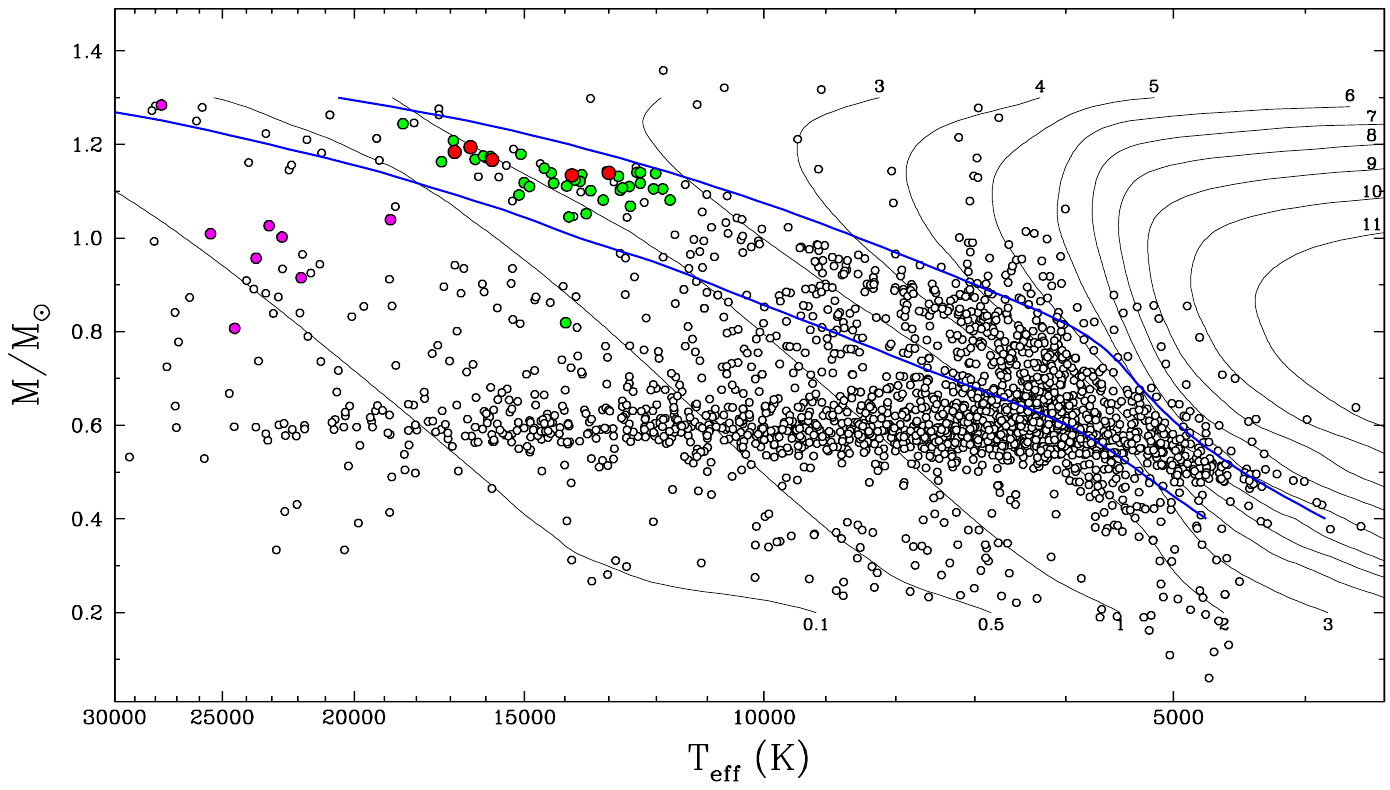}
\caption{Stellar masses as a function of effective temperature for the Montreal White Dwarf Database 100 pc sample (white dots)
along with the hot DQ (magenta) and warm DQ (green) white dwarfs. Red dots mark the 5 DAQs identified in this paper.
Solid curves are theoretical isochrones, labeled in units of Gyr, obtained from standard cooling sequences with CO-core compositions,
$q({\rm He}) \equiv M({\rm He})/M_{\star} = 10^{-2}$, and $q$(H) = $10^{-4}$. The lower blue solid curve indicates the onset
of crystallization at the center of evolving models, while the upper one indicates the locations where 80\% of the total mass has solidified.}
\label{figtm}
\end{figure*}

\citet{hollands20} proposed carbon dredge-up as the source of the DAQ phenomenon. They  suggest that the mixed hydrogen and carbon atmosphere of J0551+4135 can be explained if both hydrogen and helium masses are significantly lower than expected from single star evolution. For low helium masses, the helium layer diffuses downwards, creating a hydrogen/carbon interface, and enabling carbon dredge-up into the hydrogen envelope. Before we discuss potential evolutionary scenarios for the emergence of the DAQ subclass, we first summarize observations
relevant to this discussion.

Observations of the five DAQ white dwarfs identified in this study indicate that they display many similarities with the hot DQ and warm DQ populations: they are massive, have  carbon-enriched atmospheres, display unusual kinematics for their cooling ages, and (at least two out of five) display photometric variability due to fast rotation. 

Figure \ref{figtm} summarizes the stellar masses as a function of effective temperature for the five DAQ white dwarfs presented in this paper, along with the Montreal White Dwarf Database 100 pc sample \citep{kilic20}, as well as the hot DQ and warm DQ white dwarfs taken from the samples of \citet{coutu19}, \citet{koester19}, and Jewett et al. (in prep.). All warm DQs in this figure have been reanalyzed using the same technique as that described above, but assuming an almost undetectable trace of helium of ${\rm He/C}=1$.

We can see that hot DQ white dwarfs are prevalent above $T_{\rm eff}=18,000$ K, while the DAQ subclass emerges below about 17,000 K. The 7 hot DQs analyzed by \citet{koester19} have masses ranging from 0.81 to 1.04 \msun, and they are found outside of the CO crystallization branch. More importantly, they have masses lower than those of the DAQ white dwarfs, and hence they are unlikely progenitors to the DAQ spectral class, unless the masses of these hot DQ stars are significantly underestimated. However, J1819$-$1208 \citep{kilic23a}, the only hot DQ in the 100 pc sample, is much more massive and is also located within the crystallization sequence.

Interestingly, all DAQ and all but one of the warm DQ white dwarfs displayed in Figure \ref{figtm} lie within the crystallization boundary, and none are cooler. The only exception is the warm DQ SDSS J104052.40+063519.7 with a mass near 0.8 \msun. \citet{coutu19} note that J1040+0635 is magnetic, but further observations are needed to understand the unusual nature of this object, and to see if its relatively low inferred-mass could be due to binarity. 

Excluding J1040+0635, all of the DAQ and warm DQ stars share the same parameter space in the $M$ versus $T_{\rm eff}$ diagram, suggesting that they have a common origin.  The main difference between the DAQ and warm DQs is the amount of hydrogen they contain, as previously shown in Figure \ref{figratio}. Indeed, many of the warm DQs contain significant amounts of hydrogen \citep{dufour08,koester19}, and a few of the hot DQs as well. For example, \citet{dufour08} identified two hot DQ white dwarfs with $T_{\rm eff} \sim 21,000$ K and with visible H$\beta$ lines that indicate $\log{\rm C/H}\approx1.7$. This is similar to the C/H ratios measured in warm DQs (see Figure \ref{figratio}).

\citet{cheng19} identified a cooling anomaly of high mass white dwarfs on the crystallization sequence of CO white dwarfs \citep[the so-called Q-branch,][]{gaia18,tremblay19} and found that 5-9\% of the high-mass white dwarfs experience an extra cooling
delay of at least 8 Gyr that cannot be accounted for by standard core crystallization or merger delays. 
 
\cite{blouin21} suggested that a phase separation process involving $^{22}$Ne (not to be confused with the gravitational settling of $^{22}$Ne in the liquid phase) might generate enough energy to produce a multi-Gyr pause in the cooling of CO white dwarfs. For typical CO core compositions, it has been shown that the solid phase is depleted in $^{22}$Ne compared to the liquid. This can lead to the creation of buoyant crystals that inhibit the standard inside-out crystallization scenario. The buoyant crystals float up, and thereby displace $^{22}$Ne-rich liquid toward the center of the star in a solid--liquid distillation process that liberates copious amounts of gravitational energy \citep{isern91}. This has now been implemented in white dwarf cooling models and shown to match all the observational properties of Q-branch white dwarfs \citep{bedard24}. The exact origin of these delayed Q-branch objects remains unclear. 

The fact that the Q-branch coincides with CO crystallization and that the distillation mechanism requires CO cores suggest that these white dwarfs must harbor CO cores. This is in tension with the predictions from \citet{schwab21}, who find that for ultramassive white dwarfs with masses exceeding $1.05\,M_{\odot}$, both single-star evolution and the merger of two CO white dwarfs should yield an ONe white dwarf. However,
\citet{dominguez96} demonstrated that during the CO core formation, even a small initial rotation may have a significant impact on the
evolutionary outcome. The decrease in core pressure, brought on by the anticipated increase in angular velocities during
the compression of the CO core at the onset of the AGB phase, leads to a reduction in maximum temperature. This favors the formation of ultramassive CO white dwarfs with masses exceeding $1.05~M_{\odot}$. Furthermore, \citet{althaus21} have more recently corroborated this
phenomenon through detailed computations of single progenitor evolution, confirming the formation of ultramassive CO white dwarfs.

A possible progenitor channel for the delayed Q-branch population is the merger of white dwarfs and subgiant stars, which can produce the ultramassive CO white dwarfs with enough neutron-rich impurities needed to power the distillation mechanism \citep{shen23}. 
Given that DAQ and warm DQ stars are likely merger remnants, have old kinematic ages and are confined to the crystallization sequence (Figure~\ref{figtm}), it is natural to assume that they are currently undergoing a distillation-powered multi-Gyr cooling pause. Indeed, \citet{cheng19} find that half of the extra delayed population is made of DQ stars. Hence, we end up with the interesting idea that
the spectral evolution of DAQ and warm DQ white dwarfs may proceed at nearly constant effective temperature. We explore below the possibilities that DAQ stars evolve into warm DQ white dwarfs, or the other way around.

\subsection{Evolutionary Path of DAQs}

In the temperature range where warm DQ stars are found ($T_{\rm eff}\sim 13,000-17,000$ K), their atmospheres and stellar envelopes are strongly convective, even more so at large masses (see, e.g., Figures 9 and 10 of \citealt{rolland20}). This is certainly the case for DAQ white dwarfs as well, otherwise they would not have mixed hydrogen and carbon compositions. Hence it is not possible for hydrogen to float to the surface through ordinary diffusion, a process that requires the environment to be convectively stable. It is thus very unlikely that warm DQ/DQA white dwarfs evolve into DAQ stars.

A more likely scenario is that DAQ white dwarfs evolve into warm DQ stars through a process analogous to one of the most well-established spectral evolution scenarios, namely the DO $\rightarrow$ DA $\rightarrow$ DB/DC transformation \citep{bedard22a}. In this scenario, a hot, helium-atmosphere DO star evolves over time into a DA white dwarf when hydrogen starts floating towards the surface as a result of ordinary diffusion in a radiative envelope. In this particular context, stratified DAO stars are believed to be transitional objects. As evolution proceeds, some of these DA stars with thin enough hydrogen layers will be convectively diluted near 20,000 K and become DBA white dwarfs (objects with thicker hydrogen layers may be convectively mixed below about 12,000 K and become helium-rich DA stars or DC white dwarfs). Based on detailed calculations of the spectral evolution of white dwarfs \citep{bedard22b}, including different chemical transport mechanisms, \citet{bedard23} also predict the existence of a massive hydrogen reservoir underneath the thin superficial layer, and demonstrate that the atmospheric hydrogen in DBA stars is likely from this internal reservoir of residual hydrogen \citep[see also][]{rolland20}. This model successfully reproduces the observed H/He ratios among DBA white dwarfs, which comprise as much as 60 to 75\% of the DB population in the solar neighborhood \citep{koester15, rolland18}. Hydrogen always remains a trace element, however, in the sense that DBA white dwarfs never become DA white dwarfs at lower temperatures. 

We are likely witnessing a similar phenomenon in the DAQ and warm DQ sample, but with a very different interior, with carbon being much closer to the surface than normal post-AGB objects, likely as a result of the merger process. In this proposed scenario, a thin hydrogen atmosphere (in a hot massive DA star) is being diluted by the deeper C/He-rich convective envelope as the star cools off, gradually turning this massive DA white dwarf into a DAQ, and eventually into a warm DQA star as hydrogen is being diluted further within the deep C/He convective envelope. 

Because these massive white dwarfs are trapped on the crystallization sequence, they go through the DA to DQA spectral evolution at a relatively constant surface temperature. The amount of hydrogen in warm DQA stars is most likely too high for a simple dilution process, and a deep hydrogen reservoir may also be required, as in DBA white dwarfs \citep{rolland20,bedard23}. 

In this scenario, the immediate progenitors of DAQ stars are probably massive DA white dwarfs with extremely thin hydrogen layers, although carbon enrichment seems to be a rare phenomenon among the relatively young, ultramassive DA white dwarf population. However, it is also possible that the even more distant progenitors are massive, hot DQ stars such as J1819$-$1208 (see Figure \ref{figtm}), in which residual hydrogen present in the radiative envelope eventually floats to the surface through ordinary diffusion. If the total hydrogen mass is too small, these hot DQ stars may evolve directly into warm DQs or DQAs, without becoming DAQ white dwarfs. A similar situation occurs in the context of DB white dwarfs, where hot DO stars may evolve directly into DB or DBA white dwarfs, without ever becoming DA stars, if the total mass of hydrogen is too small \citep{bedard23}. \citet{althaus05} highlighted the impact of the internal diffusion of diluted hydrogen during the born-again phase on the
formation of white dwarfs, and demonstrated the evolutionary connection between PG1159-DB-DC/DQ white dwarfs.

To summarize, we propose two potential evolutionary channels involving warm DQ and DAQ white dwarfs. Depending on how much hydrogen is present in the stellar envelope, the white dwarf merger remnant may appear as a massive DA or a hot DQ. The proposed channels are:

1) DA $\rightarrow$ DAQ $\rightarrow$ warm DQA.   

2) Hot DQ(A) $\rightarrow$ warm DQ(A).

\noindent In the second scenario, hot DQs may evolve into warm DQs where hydrogen is either not seen or has trace amounts in the atmosphere. 

In both of our proposed evolutionary scenarios, the stars are trapped on the crystallization sequence. This would explain why we do not observe their cooler counterparts (massive cool DQ white dwarfs) in the solar neighborhood (see, e.g., \citealt{kilic20}, \citealt{caron23}). Detailed evolutionary calculations for the spectral evolution of hot and warm DQs are currently not available, but would be helpful in further understanding the emergence of the DAQ subclass.

\section{Conclusions}

Through follow-up spectroscopy of massive white dwarf candidates within 100 pc, we identified two new DAQ white dwarfs with 
mixed carbon and hydrogen atmospheres. In addition, based on a detailed model atmosphere analysis, we demonstrated that two additional
DAQ white dwarfs were overlooked in the literature. These four objects increase the sample of DAQ white dwarfs from one
\citep{hollands20} to five. 

The DAQ white dwarf sample shows several characteristics that favor a merger origin. All five stars
have masses in the range 1.14 to 1.19 \msun, roughly twice the average mass for the DA white dwarfs in the solar
neighborhood. In addition, all five are relatively young with cooling ages of about 1 Gyr, but kinematically
old, with tangential velocities greater than 50 km s$^{-1}$, and as high as  133 km s$^{-1}$. Their Galactic UVW velocities
point to a thick disk or halo origin. In addition, we detect rapid rotation in at least two of these objects. These characteristics are
all very similar to the warm DQ white dwarf population, which also lie within the crystallization sequence.

Given the similarities between the warm DQ population, and the DAQ white dwarfs discovered here, we propose two potential
evolutionary channels for DAQ and warm DQ stars. 
The DAQ population likely emerges as massive DA white dwarfs, produced in white dwarf mergers, are convectively mixed. 
\citet{shen23} propose subgiant + CO white dwarf
mergers as the progenitors of a significant fraction of the Q-branch white dwarfs with delayed cooling times. Further theoretical studies
of the spectral evolution of merger products, including hot and warm DQs and the DAQs would be beneficial for understanding the emergence
of the DAQ subclass.

\begin{acknowledgements}

We are grateful to Antoine B\'edard for useful discussions. This work is supported in part by the NSF under grant  AST-2205736, the NASA under grants 80NSSC22K0479, 80NSSC24K0380, and 80NSSC24K0436, the NSERC Canada, the Fund FRQ-NT (Qu\'ebec), the Canadian Institute for Theoretical Astrophysics (CITA) National Fellowship Program, and by the Smithsonian Institution.

Based on observations obtained at the MMT Observatory, a joint facility of the Smithsonian  Institution and the University of Arizona.

The Apache Point Observatory 3.5-meter telescope is owned and operated by the Astrophysical Research Consortium.

\end{acknowledgements}

\facilities{MMT (Blue Channel spectrograph), APO 3.5m (KOSMOS spectrograph, ARCTIC imager)}



\begin{thebibliography}{}
\expandafter\ifx\csname natexlab\endcsname\relax\def\natexlab#1{#1}\fi

\bibitem[Althaus et al.(2021)]{althaus21} Althaus, L.~G., Gil-Pons, P., C{\'o}rsico, A.~H., et al.\ 2021, \aap, 646, A30. doi:10.1051/0004-6361/202038930

\bibitem[Althaus et al.(2005)]{althaus05} Althaus, L.~G., Serenelli, A.~M., Panei, J.~A., et al.\ 2005, \aap, 435, 631. doi:10.1051/0004-6361:20041965

\bibitem[Bailer-Jones et al.(2021)]{bailer21} Bailer-Jones, C.~A.~L., Rybizki, J., Fouesneau, M., et al.\ 2021, \aj, 161, 147.

\bibitem[B{\'e}dard et al.(2024)]{bedard24} B{\'e}dard, A., Blouin, S., and Cheng, S.\ 2024, Nature, https://doi.org/10.1038/s41586-024-07102-y

\bibitem[{{B{\'e}dard} {et~al.}(2023){B{\'e}dard}, {Bergeron}, \& {Brassard}}]{bedard23} {B{\'e}dard}, A., {Bergeron}, P., \& {Brassard}, P. 2023, \apj, 946, 24

\bibitem[B{\'e}dard et al.(2022a)]{bedard22a} B{\'e}dard, A., Bergeron, P., \& Brassard, P.\ 2022a, \apj, 930, 8.

\bibitem[B{\'e}dard et al.(2022b)]{bedard22b} B{\'e}dard, A., Brassard, P., Bergeron, P., et al.\ 2022b, \apj, 927, 128. 

\bibitem[B{\'e}dard et al.(2020)]{bedard20} B{\'e}dard, A., Bergeron, P., Brassard, P., et al.\ 2020, \apj, 901, 93. 

\bibitem[{{Bergeron} {et~al.}(2019){Bergeron}, {Dufour}, {Fontaine}, {Coutu},
  {Blouin}, {Genest-Beaulieu}, {B{\'e}dard}, \& {Rolland }}]{bergeron19}
{Bergeron}, P., {Dufour}, P., {Fontaine}, G., {et~al.} 2019, \apj, 876, 67

\bibitem[{{Bianchi} {et~al.}(2017){Bianchi}, {Shiao}, \& {Thilker}}]{bianchi17}
{Bianchi}, L., {Shiao}, B., \& {Thilker}, D. 2017, \apjs, 230, 24

\bibitem[{{Blouin} {et~al.}(2021){Blouin}, {Daligault}, \& {Saumon}}]{blouin21}
{Blouin}, S., {Daligault}, J., \& {Saumon}, D. 2021, \apjl, 911, L5

\bibitem[{{Blouin} {et~al.}(2019){Blouin}, {Dufour}, {Thibeault}, \&
  {Allard}}]{blouin19}
{Blouin}, S., {Dufour}, P., {Thibeault}, C., \& {Allard}, N.~F. 2019, \apj,
  878, 63

\bibitem[{{Caiazzo} {et~al.}(2021){Caiazzo}, {Burdge}, {Fuller}, {Heyl},
  {Kulkarni}, {Prince}, {Richer}, {Schwab}, {Andreoni}, {Bellm}, {Drake},
  {Duev}, {Graham}, {Helou}, {Mahabal}, {Masci}, {Smith}, \&
  {Soumagnac}}]{caiazzo21}
{Caiazzo}, I., {Burdge}, K.~B., {Fuller}, J., {et~al.} 2021, Nature, 595, 39

\bibitem[Capitanio et al.(2017)]{capitanio17} Capitanio, L., Lallement, R., Vergely, J.~L., et al.\ 2017, \aap, 606, A65. 

\bibitem[Caron et al.(2023)]{caron23} Caron, A., Bergeron, P., Blouin, S., et al.\ 2023, \mnras, 519, 4529. 

\bibitem[{{Cheng} {et~al.}(2019){Cheng}, {Cummings}, \& {M{\'e}nard}}]{cheng19}
{Cheng}, S., {Cummings}, J.~D., \& {M{\'e}nard}, B. 2019, \apj, 886, 100

\bibitem[{{Chiba} \& {Beers}(2000)}]{chiba00}
{Chiba}, M., \& {Beers}, T.~C. 2000, \aj, 119, 2843

\bibitem[{{Clayton} {et~al.}(2007){Clayton}, {Geballe}, {Herwig}, {Fryer}, \&
  {Asplund}}]{clayton07}
{Clayton}, G.~C., {Geballe}, T.~R., {Herwig}, F., {Fryer}, C., \& {Asplund}, M.
  2007, \apj, 662, 1220

\bibitem[{{Coutu} {et~al.}(2019){Coutu}, {Dufour}, {Bergeron}, {Blouin},
  {Loranger}, {Allard}, \& {Dunlap}}]{coutu19}
{Coutu}, S., {Dufour}, P., {Bergeron}, P., {et~al.} 2019, \apj, 885, 74

\bibitem[Dominguez et al.(1996)]{dominguez96} Dominguez, I., Straniero, O., Tornambe, A., et al.\ 1996, \apj, 472, 783. doi:10.1086/178106

\bibitem[{{Dufour} {et~al.}(2017){Dufour}, {Blouin}, {Coutu},
  {Fortin-Archambault}, {Thibeault}, {Bergeron}, \& {Fontaine}}]{dufour17}
{Dufour}, P., {Blouin}, S., {Coutu}, S., {et~al.} 2017, in Astronomical Society
  of the Pacific Conference Series, Vol. 509, 20th European White Dwarf
  Workshop, ed. P.~E. {Tremblay}, B.~{Gaensicke}, \& T.~{Marsh}, 3

\bibitem[{{Dufour} {et~al.}(2008){Dufour}, {Fontaine}, {Liebert}, {Schmidt}, \&
  {Behara}}]{dufour08}
{Dufour}, P., {Fontaine}, G., {Liebert}, J., {Schmidt}, G.~D., \& {Behara}, N.
  2008, \apj, 683, 978

\bibitem[{{Dunlap} \& {Clemens}(2015)}]{dunlap15}
{Dunlap}, B.~H., \& {Clemens}, J.~C. 2015, in Astronomical Society of the
  Pacific Conference Series, Vol. 493, 19th European Workshop on White Dwarfs,
  ed. P.~{Dufour}, P.~{Bergeron}, \& G.~{Fontaine}, 547

\bibitem[Gaia Collaboration et al.(2018)]{gaia18} Gaia Collaboration, Babusiaux, C., van Leeuwen, F., et al.\ 2018, \aap, 616, A10.

\bibitem[{{Greenstein}(1984)}]{greenstein84}
{Greenstein}, J.~L. 1984, \apj, 276, 602

\bibitem[{{Heber}(2009)}]{heber09}
{Heber}, U. 2009, \araa, 47, 211

\bibitem[{{Hermes} {et~al.}(2017){Hermes}, {G{\"a}nsicke}, {Kawaler}, {Greiss},
  {Tremblay}, {Gentile Fusillo}, {Raddi}, {Fanale}, {Bell}, {Dennihy}, {Fuchs},
  {Dunlap}, {Clemens}, {Montgomery}, {Winget}, {Chote}, {Marsh}, \&
  {Redfield}}]{hermes17b}
{Hermes}, J.~J., {G{\"a}nsicke}, B.~T., {Kawaler}, S.~D., {et~al.} 2017, \apjs,
  232, 23

\bibitem[{{Hollands} {et~al.}(2020){Hollands}, {Tremblay}, {G{\"a}nsicke},
  {Camisassa}, {Koester}, {Aungwerojwit}, {Chote}, {C{\'o}rsico}, {Dhillon},
  {Gentile-Fusillo}, {Hoskin}, {Izquierdo}, {Marsh}, \& {Steeghs}}]{hollands20}
{Hollands}, M.~A., {Tremblay}, P.~E., {G{\"a}nsicke}, B.~T., {et~al.} 2020,
  Nature Astronomy, 4, 663

\bibitem[{{Iben} \& {Tutukov}(1984)}]{iben84}
{Iben}, I., J., \& {Tutukov}, A.~V. 1984, \apjs, 54, 335

\bibitem[Isern et al.(1991)]{isern91} Isern, J., Hernanz, M., Mochkovitch, R., et al.\ 1991, \aap, 241, L29

\bibitem[{{Kawaler}(2015)}]{kawaler15}
{Kawaler}, S.~D. 2015, in Astronomical Society of the Pacific Conference
  Series, Vol. 493, 19th European Workshop on White Dwarfs, ed. P.~{Dufour},
  P.~{Bergeron}, \& G.~{Fontaine}, 65

\bibitem[{{Kawka}(2020)}]{kawka20}
Kawka, A.\ 2020, White Dwarfs as Probes of Fundamental Physics: Tracers of Planetary, Stellar and Galactic Evolution, 357, 60. 

\bibitem[{{Kawka} {et~al.}(2023){Kawka}, {Ferrario}, \& {Vennes}}]{kawka23}
{Kawka}, A., {Ferrario}, L., \& {Vennes}, S. 2023, \mnras, 520, 6299

\bibitem[{{Kepler} {et~al.}(2015){Kepler}, {Pelisoli}, {Koester}, {Ourique},
  {Kleinman}, {Romero}, {Nitta}, {Eisenstein}, {Costa}, {K{\"u}lebi}, {Jordan},
  {Dufour}, {Giommi}, \& {Rebassa-Mansergas}}]{kepler15}
{Kepler}, S.~O., {Pelisoli}, I., {Koester}, D., {et~al.} 2015, \mnras, 446,
  4078

\bibitem[{{Kilic} {et~al.}(2020){Kilic}, {Bergeron}, {Kosakowski}, {Brown},
  {Ag{\"u}eros}, \& {Blouin}}]{kilic20}
{Kilic}, M., {Bergeron}, P., {Kosakowski}, A., {et~al.} 2020, \apj, 898, 84

\bibitem[{{Kilic} {et~al.}(2023{\natexlab{a}}){Kilic}, {C{\'o}rsico}, {Moss},
  {Jewett}, {De Ger{\'o}nimo}, \& {Althaus}}]{kilic23b}
{Kilic}, M., {C{\'o}rsico}, A.~H., {Moss}, A.~G., {et~al.} 2023{\natexlab{a}},
  \mnras, 522, 2181

\bibitem[{{Kilic} {et~al.}(2021){Kilic}, {Kosakowski}, {Moss}, {Bergeron}, \&
  {Conly}}]{kilic21b}
{Kilic}, M., {Kosakowski}, A., {Moss}, A.~G., {Bergeron}, P., \& {Conly}, A.~A.
  2021, \apjl, 923, L6

\bibitem[{{Kilic} {et~al.}(2023{\natexlab{b}}){Kilic}, {Moss}, {Kosakowski},
  {Bergeron}, {Conly}, {Brown}, {Toonen}, {Williams}, \& {Dufour}}]{kilic23a}
{Kilic}, M., {Moss}, A.~G., {Kosakowski}, A., {et~al.} 2023{\natexlab{b}},
  \mnras, 518, 2341

\bibitem[{{Koester} \& {Kepler}(2015)}]{koester15}
{Koester}, D., \& {Kepler}, S.~O. 2015, \aap, 583, A86

\bibitem[{{Koester} \& {Kepler}(2019)}]{koester19}
---. 2019, \aap, 628, A102

\bibitem[{{Liebert}(1983)}]{liebert83}
{Liebert}, J. 1983, \pasp, 95, 878

\bibitem[{{Lor{\'e}n-Aguilar} {et~al.}(2009){Lor{\'e}n-Aguilar}, {Isern}, \&
  {Garc{\'\i}a-Berro}}]{loren09}
{Lor{\'e}n-Aguilar}, P., {Isern}, J., \& {Garc{\'\i}a-Berro}, E. 2009, \aap,
  500, 1193

\bibitem[{{McCleery} {et~al.}(2020){McCleery}, {Tremblay}, {Gentile Fusillo},
  {Hollands}, {G{\"a}nsicke}, {Izquierdo}, {Toonen}, {Cunningham}, \&
  {Rebassa-Mansergas}}]{mccleery20}
{McCleery}, J., {Tremblay}, P.-E., {Gentile Fusillo}, N.~P., {et~al.} 2020,
  \mnras, 499, 1890

\bibitem[{{Moss} {et~al.}(2023){Moss}, {Kilic}, {Bergeron}, {Firgard}, \&
  {Brown}}]{moss23}
{Moss}, A., {Kilic}, M., {Bergeron}, P., {Firgard}, M., \& {Brown}, W. 2023,
  \mnras, 523, 5598

\bibitem[{{Nomoto} \& {Iben}(1985)}]{nomoto85}
{Nomoto}, K., \& {Iben}, I., J. 1985, \apj, 297, 531

\bibitem[{{Pelletier} {et~al.}(1986){Pelletier}, {Fontaine}, {Wesemael},
  {Michaud}, \& {Wegner}}]{pelletier86}
{Pelletier}, C., {Fontaine}, G., {Wesemael}, F., {Michaud}, G., \& {Wegner}, G.
  1986, \apj, 307, 242

\bibitem[{{Press} {et~al.}(1986){Press}, {Flannery}, \& {Teukolsky}}]{press86}
{Press}, W.~H., {Flannery}, B.~P., \& {Teukolsky}, S.~A. 1986, {Numerical
  recipes. The art of scientific computing}

\bibitem[{{Pshirkov} {et~al.}(2020){Pshirkov}, {Dodin}, {Belinski},
  {Zheltoukhov}, {Fedoteva}, {Voziakova}, {Potanin}, {Blinnikov}, \&
  {Postnov}}]{pshirkov20}
{Pshirkov}, M.~S., {Dodin}, A.~V., {Belinski}, A.~A., {et~al.} 2020, \mnras,
  499, L21

\bibitem[{{Rolland} {et~al.}(2018){Rolland}, {Bergeron}, \&
  {Fontaine}}]{rolland18}
{Rolland}, B., {Bergeron}, P., \& {Fontaine}, G. 2018, \apj, 857, 56

\bibitem[{{Rolland} {et~al.}(2020){Rolland}, {Bergeron}, \&
  {Fontaine}}]{rolland20}
---. 2020, \apj, 889, 87

\bibitem[{{Schmidt} {et~al.}(1989){Schmidt}, {Weymann}, \& {Foltz}}]{schmidt89}
{Schmidt}, G.~D., {Weymann}, R.~J., \& {Foltz}, C.~B. 1989, \pasp, 101, 713

\bibitem[Sch{\"o}nrich et al.(2010)]{schonrich10} Sch{\"o}nrich, R., Binney, J., \& Dehnen, W.\ 2010, \mnras, 403, 1829.

\bibitem[{{Schwab}(2021)}]{schwab21}
{Schwab}, J. 2021, \apj, 906, 53

\bibitem[Shen et al.(2023)]{shen23} Shen, K.~J., Blouin, S., \& Breivik, K.\ 2023, \apjl, 955, L33.

\bibitem[{{Shen} {et~al.}(2012){Shen}, {Bildsten}, {Kasen}, \&
  {Quataert}}]{shen12}
{Shen}, K.~J., {Bildsten}, L., {Kasen}, D., \& {Quataert}, E. 2012, \apj, 748,
  35

\bibitem[{{Sion} {et~al.}(1983){Sion}, {Greenstein}, {Landstreet}, {Liebert},
  {Shipman}, \& {Wegner}}]{sion83}
{Sion}, E.~M., {Greenstein}, J.~L., {Landstreet}, J.~D., {et~al.} 1983, \apj,
  269, 253

\bibitem[{{Thejll} {et~al.}(1990){Thejll}, {Shipman}, {MacDonald}, \&
  {Macfarland}}]{thejll90}
{Thejll}, P., {Shipman}, H.~L., {MacDonald}, J., \& {Macfarland}, W.~M. 1990,
  \apj, 361, 197

\bibitem[{{Toonen} {et~al.}(2012){Toonen}, {Nelemans}, \& {Portegies
  Zwart}}]{toonen12}
{Toonen}, S., {Nelemans}, G., \& {Portegies Zwart}, S. 2012, \aap, 546, A70

\bibitem[{{Tremblay} {et~al.}(2019){Tremblay}, {Fontaine}, {Fusillo}, {Dunlap},
  {G{\"a}nsicke}, {Hollands}, {Hermes}, {Marsh}, {Cukanovaite}, \&
  {Cunningham}}]{tremblay19}
{Tremblay}, P.-E., {Fontaine}, G., {Fusillo}, N. P.~G., {et~al.} 2019, \nat,
  565, 202

\bibitem[{{Vincent} {et~al.}(2020){Vincent}, {Bergeron}, \&
  {Lafreni{\`e}re}}]{vincent20}
{Vincent}, O., {Bergeron}, P., \& {Lafreni{\`e}re}, D. 2020, \aj, 160, 252

\bibitem[{{Webbink}(1984)}]{webbink84}
{Webbink}, R.~F. 1984, \apj, 277, 355

\bibitem[{{Wegner} \& {Koester}(1985)}]{wegner85}
{Wegner}, G., \& {Koester}, D. 1985, \apj, 288, 746

\bibitem[{{Williams} {et~al.}(2022){Williams}, {Hermes}, \&
  {Vanderbosch}}]{williams22}
{Williams}, K.~A., {Hermes}, J.~J., \& {Vanderbosch}, Z.~P. 2022, \aj, 164, 131

\bibitem[{{Yoon} {et~al.}(2007){Yoon}, {Podsiadlowski}, \& {Rosswog}}]{yoon07}
{Yoon}, S.~C., {Podsiadlowski}, P., \& {Rosswog}, S. 2007, \mnras, 380, 933

\bibitem[{{Zuckerman} {et~al.}(2007){Zuckerman}, {Koester}, {Melis}, {Hansen},
  \& {Jura}}]{zuckerman07}
{Zuckerman}, B., {Koester}, D., {Melis}, C., {Hansen}, B.~M., \& {Jura}, M.
  2007, \apj, 671, 872

\end{thebibliography}

\end{document}